\documentclass{article}
\usepackage{arxiv}
\usepackage[T1]{fontenc}    
\usepackage{setspace}
\usepackage{hyperref}       
\usepackage{url}            
\usepackage{booktabs}       
\usepackage{amsfonts}       
\usepackage{nicefrac}       
\usepackage{microtype}      
\usepackage{lipsum}		
\usepackage{graphicx}
\usepackage{natbib}
\usepackage{doi}
\usepackage{amsthm,amsmath,amssymb}
\usepackage{graphicx,bm}
\usepackage{color}
\usepackage{xcolor}
\usepackage{lscape}
\usepackage{rotating}
\usepackage{subcaption}
\usepackage{algorithm}
\usepackage{algpseudocode}

\newcommand{\bs}[1] {\boldsymbol{#1}}

\title{Supervised and Unsupervised Mapping of Binary Variables: A proximity perspective}

\author{ \href{https://orcid.org/0000-0001-7308-6210}{\includegraphics[scale=0.06]{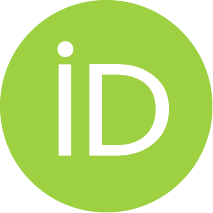}\hspace{1mm}Mark de Rooij}\\
	Methodology and Statistics department\\
	Leiden University\\
	Leiden, The Netherlands \\
	\texttt{rooijm@fsw.leidenuniv.nl} \\
	\And
	\href{https://orcid.org/0000-0001-6143-1101}{\includegraphics[scale=0.06]{orcid.pdf}\hspace{1mm}Dion Woestenburg} \\
	Methodology and Statistics department\\
	Leiden University\\
	Leiden, The Netherlands \\
	\texttt{d.h.a.woestenburg@fsw.leidenuniv.nl} \\	
	\And
	\href{https://orcid.org/0000-0002-8062-538x}{\includegraphics[scale=0.06]{orcid.pdf}\hspace{1mm}Frank Busing} \\
	Methodology and Statistics department\\
	Leiden University\\
	Leiden, The Netherlands \\
	\texttt{busing@fsw.leidenuniv.nl} \\	
}

\date{}

\renewcommand{\headeright}{}
\renewcommand{\undertitle}{}
\renewcommand{\shorttitle}{Proximity Mapping of Multivariate Binary Data}

\begin{document}
\maketitle


\begin{abstract}
We propose a new mapping tool for supervised and unsupervised analysis of multivariate binary data with multiple items, questions, or response variables. The mapping assumes an underlying proximity response function, where participants can have multiple reasons to disagree or say ``no'' to a question. The probability to endorse, or to agree with an item depends on an item specific parameter and the distance in a joint space between a point representing the item and a point representing the participant. The item specific parameter defines a circle in the joint space around the location of the item such that for participants positioned within the circle the endorsement probability is larger than 0.5. For map estimation, we develop and test an MM-algorithm in which the negative log-likelihood function is majorized with a weighted least squares function. The weighted least squares function can be minimized with standard algorithms for multidimensional unfolding. To illustrate the new mapping, two empirical data sets are analyzed. The mappings are interpreted in detail and the unsupervised map is compared to a visualization based on correspondence analysis. In a Monte Carlo study, we test the performance of the algorithm in terms of recovery of population parameters and conclude that this recovery is adequate. A second Monte Carlo study investigates the predictive performance of the new mapping compared to a similar mapping with a monotone response function. 
\end{abstract}


\newpage 
\section{Introduction}

Multivariate binary data are often collected in different fields of research. In such investigations, for a set of participants,  dichotomous responses are collected on a set of response variables or items. In different sciences, mapping might be considered an important tool. For example, researchers in political science are interested in vote intentions for an upcoming election. In the Dutch parliamentary election studies \citep{dpes0203}, for example, participants are asked to indicate which political parties were still under consideration for a vote a few months prior to the election. In psychiatry, researchers are interested in mental disorders, such as depression and anxiety and comorbidity of such disorders. Mental disorders are highly prevalent in modern western societies and a high degree of comorbidity can often be observed among these disorders. In the Netherlands Study for Depression and Anxiety \citep[NESDA,][] {penninx2008netherlands} data were collected on a large number of participants on different mental disorders \citep{spinhoven2009role}. In health sciences, researchers are interested in drug consumption profiles, that is, which people use the same set of drugs. \cite{fehrman2017five} are interested in subjects' drug consumption profiles and collected data about the consumption of 18 different drugs. As a final example, \cite{sugiyama1975} collected data from 4,243 Japanese persons, where each participant had to pick any of six different religious practices, leading to binary variables on these six items.
 
In some studies, besides these binary response variables also characteristics of the persons are available. In the NESDA study and in \cite{fehrman2017five}, personality characteristics of the participants are available and the researchers are interested in how these personality characteristics influence mental disorders or drug use. In the Dutch parliamentary election studies, opinions on several topics are available for the participants and researchers are interested in the link between the opinions and the vote intentions. Without such predictor variables the analysis is \emph{unsupervised} while the analysis is \emph{supervised} with such predictors. 


For the analysis of binary data, it is important to distinguish between two types of response processes \citep{thurstone1929measurement, coombs1964theory, polak2011}. In a unipolar scale or map, item responses are monotonically related to the position of the person on the map. The items are so-called \emph{dominance items}. Mathematical problems are a typical example of dominance items where subjects with a higher mathematical ability have a higher probability of getting the problem right. For dominance items the subjects are partitioned into two homogeneous group, that is, both the group of subjects who answer the item correct (1) and the group who answer the item wrong (0) constitute homogeneous groups. 

In a bipolar scale or map the item responses are characterized by the proximity between the item and the respondent: The item responses are functions of the distance between the position of an item and the position of a person.  The items are so-called \emph{proximity items}. For proximity items, only the respondents who answer yes (or 1) form a homogeneous group. The respondents who answer no, might do so because of a diverse set of reasons. 

In classical multivariate analysis, principal component analysis \citep[PCA, ][]{pearson1901principal, hotelling1936simplified, jolliffe2002principal} is the standard tool for the analysis of dominance items whereas multidimensional unfolding \citep[MDU, ][]{coombs1964theory, heiser1981unfolding, busing2010advances} and Correspondence Analysis \citep[CA, ][]{heiser1981unfolding, terbraak1985correspondence, polak2009two} are the standard tools for the analysis of proximity processes. PCA, MDU, and CA provide low-dimensional geometric mappings of the data. 

For binary data a logistic framework is most natural, for both supervised and unsupervised analysis. For the analysis of a single binary response variable and a set of predictors logistic regression is preferred over linear regression. Let us define the probability that person \(i\) answers yes (1) for response variable \(r\) by \(\pi_{ir} = P(Y_{ir} = 1)\), where $Y_{ir}$ is the response variable. In logistic models, these probabilities are defined by the function
\[
\pi_{ir} = \frac{\exp(\theta_{ir})}{1 + \exp(\theta_{ir})} = \frac{1}{1 + \exp(-\theta_{ir})},
\] 
where $\theta_{ir}$ is the canonical log-odds form or, in generalized linear model terms, the linear predictor. 

For unsupervised analysis of a binary data matrix within the logistic framework, \cite{deleeuw2006principal} developed logistic PCA, 
where \(\theta_{ir}\) is defined in geometric terms as \(\theta_{ir} = m_r + \langle \mathbf{u}_{i}, \mathbf{v}_{r} \rangle\)
with \(\langle \cdot, \cdot \rangle\) the inner product of the \emph{person scores} ($\mathbf{u}_i$) and \emph{factor loadings} ($\mathbf{v}_r$)  and $m_r$ an offset for item $r$. This geometric representation gives a dominance perspective on the data. \cite{deleeuw2006principal}, generalizing earlier work of \cite{groenen2003}, developed a majorization-minimization (MM) algorithm that transforms the likelihood problem into an iterative least squares problem, such that in every iteration a PCA of \emph{working responses} has to be solved, which can be computed using the singular value decomposition. 

Instead of the dominance perspective (i.e., the inner product representation) we can also use a proximity perspective with a distance representation. In that case, we represent the observations and response variables with points in an $S$-dimensional Euclidean space. The coordinates of the person points are given by the vectors $\mathbf{u}_i$ and those of the items by $\mathbf{v}_r$. The Euclidean distance between these two points ($d(\mathbf{u}_{i}, \mathbf{v}_{r})$) represents, in an inverse manner, the probability for a "yes" or 1 on response variable $r$ for observation $i$: the smaller the distance the larger the probability. This two-mode distance function is central in Multidimensional Unfolding, a generalization of Multidimensional Scaling to rectangular proximity data and often used in the analysis of preference data. 

When predictors ($\mathbf{x}_i$) are available we can perform a supervised analysis and the vector $\mathbf{u}_i$ can be constrained to be a function of these predictors, that is, $\mathbf{u}_i = \mathbf{B}^\prime\mathbf{x}_i$ with $\mathbf{B}$ a matrix of regression coefficients. In the context of dominance items this leads to a logistic reduced-rank regression model \citep{yee2003reduced, derooij2023new}. \cite{derooij2023new} generalized the MM-algorithm of De Leeuw for this supervised case, where again in each iteration of the MM-algorithm a least squares problem needs to be solved, that is, updates are given by a generalized singular value decomposition of \emph{working responses}. For proximity items, we can apply a similar constraint. The person points are constrained to be linear functions of the predictor variables. Instead of multidimensional unfolding, we need restricted multidimensional unfolding to incorporate these constraints \citep{busing2010restricted}. 


In this paper, we will develop and investigate a logistic mapping based on this proximity perspective using two-mode distances, with and without restrictions. Specifically, we define \(\theta_{ir} = m_r - d(\mathbf{u}_{i}, \mathbf{v}_{r}) \), where either the vectors $\mathbf{u}_i$ are estimated freely (unsupervised) or constrained to be functions of the predictors \(\mathbf{x}_{i}\) (supervised). We will show that such a mapping allows for a larger number of response profiles compared to a mapping based on the dominance perspective. We develop an MM algorithm for estimation of the map.  


This paper is organized as follows. In the next Section, we show our mapping in detail and develop an MM-algorithm for computing the map. We also discuss model selection and assessment issues. In Section \ref{sec:applications}, we analyze Sugiyama's religious practices data and the Dutch election data. For the religious data, we compare our unsupervised solution with a correspondence analysis solution as shown in \cite{heiser1981unfolding}. For the Dutch election data we first show an unsupervised analysis followed by a supervised analysis. In Section \ref{sec:sims}, we discuss Monte Carlo studies investigating the parameter recovery of the unsupervised and supervised algorithms and the predictive performance of the supervised mapping. We conclude the paper with a discussion. 

\section{The Mapping}
\subsection{Multivariate Binary data}

We have a data set \(\{\mathbf{y}_i\}_{i=1}^{n}\) where
\(\mathbf{y}_i \in \{0,1\}^R\), with $R$ the number of response variables. Based on these observations, we define \(\mathbf{q}_i = 2\mathbf{y}_i - 1\), such that $q_{ir}$ is either 1 or -1. 
When characteristics of the participants are available, we collect them in \(\{\mathbf{x}_i\}_{i=1}^{n}\), with \(\mathbf{x}_i \in {\rm I\!R}^P\) where $P$ denotes the number of predictor variables. 

\subsection{Probabilities}\label{sec:probs}

We impose a geometric structure on the probabilities \(\pi_{ir} = Pr(Y_{ir} = 1)\) for $r = 1,\ldots,R$ as
\[
\pi_{ir} = \frac{\exp(m_r - d(\mathbf{u}_i, \mathbf{v}_r))}{1 + \exp(m_r - d(\mathbf{u}_i, \mathbf{v}_r))} = \frac{1}{1 + \exp(d(\mathbf{u}_i, \mathbf{v}_r) - m_r)},
\] 
where \(d(\mathbf{u}_i, \mathbf{v}_r)\) is the two-mode Euclidean
distance 
\[
d(\mathbf{u}_i, \mathbf{v}_r) = \sqrt{\sum_{s=1}^S (u_{is} - v_{rs})^2}
\] 
in pre-chosen dimensionality \(S\). The \(S\)-vectors
\(\mathbf{u}_i\) and \(\mathbf{v}_r\) denote the coordinates of points
representing the persons and items in the Euclidean
space. The parameter \(m_r\) is related to the response variables and 
determines the maximum height of the probabilities, that is, when
\(m_r = 0\), for example, the probabilities cannot exceed 0.5. 

\begin{figure}
\begin{center}
\includegraphics[width = .7\textwidth]{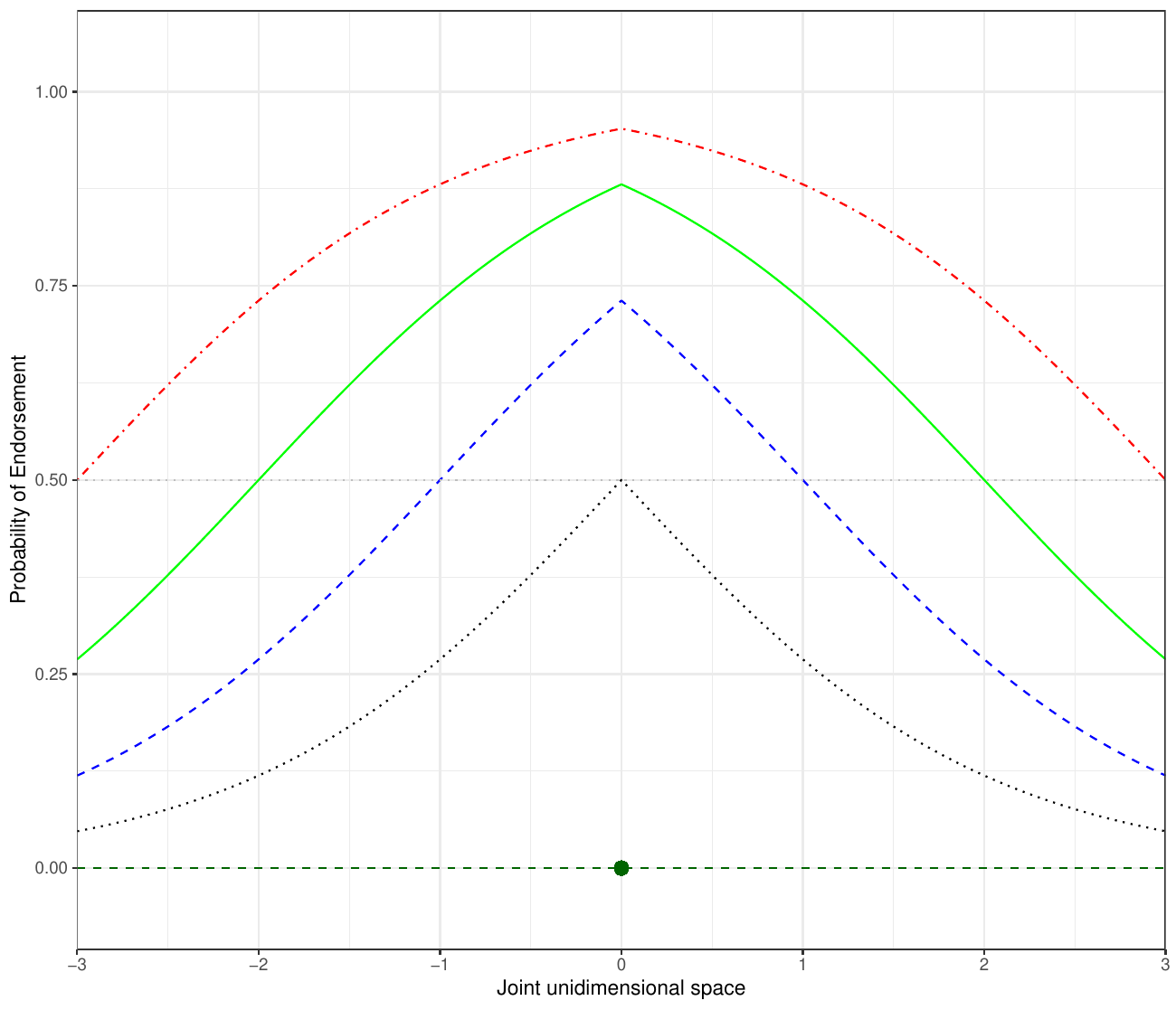}
\caption{Probability of endorsement for subjects on the real line and an item located at 0, for different values of $m_r$. The dotted black curve is for $m_r = 0$, dashed blue for $m_r = 1$, solid green for $m_r = 2$ and dashed-dotted red for $m_r = 3$.}
\label{fig:probs}
\end{center}
\end{figure}

It is instructive to see how the $m_r$ parameter influences the probabilities. Therefore, Figure \ref{fig:probs} shows the probabilities for a joint unidimensional space with a single item located at the origin ($v = 0$) and persons on the real line ($u \in [-3, 3]$) for different values of $m_r$, that is 0 (dotted black), 1 (dashed blue), 2 (solid green), and 3 (dashed-dotted red). We see that the probabilities peak at the position of the item, that is, when the distance between item and subject equals zero. Larger \(m_r\) lead to higher probabilities, but also to wider \emph{regions of endorsement}, that are, regions in the joint space where the probabilities for participants are larger than 0.5. More precisely, when 
\(d(\mathbf{u}_i, \mathbf{v}_r) = m_r\) the probability equals 0.5. Consequently, when $m_r < 0$ all probabilities are smaller than 0.5. When $m_r > 0$ there are two points on the joint scale where the probability equals 0.5, one at a distance of $m_r$ from the item to the left and one at the same distance to the right. From Figure \ref{fig:probs} it also becomes clear that the group of participants that does not endorse the item is heterogeneous, that is, persons at both extremes have low probabilities. 

\begin{figure}
\begin{center}
\includegraphics[width = .7\textwidth]{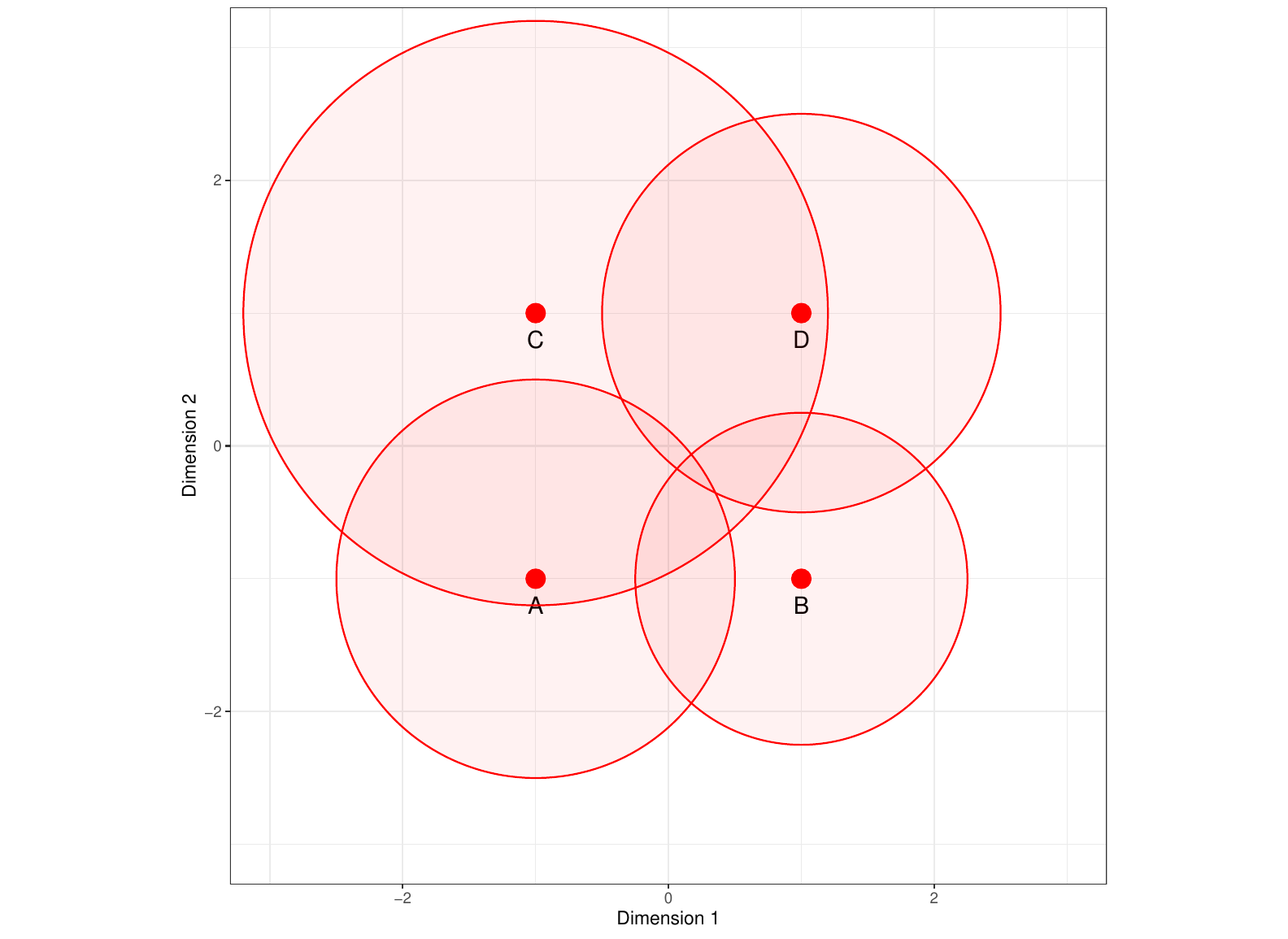}
\caption{A two-dimensional mapping with four items or response variables. The four points indicate the positions of the items, while the circles indicate the regions of endorsement with $P(Y_{ir}) \geq 0.5$, that are regions where participants have a probability larger than 0.5 to endorse the specific item.}
\label{fig:2d}
\end{center}
\end{figure}

In two-dimensional solutions, the regions of endorsement become circles centered at the item point and with radius equal to $m_r$. When a person point falls within such a circle the probability of endorsement is larger than 0.5 and we classify the subject for that response variable in the class responding 1, whereas if the person point falls outside the circle we classify it in the class responding with 0. Persons may fall outside the circle in any direction, showing the heterogeneity of these participants. An example configuration is shown in Figure \ref{fig:2d}, where we show four response variables or items ($R = 4$). Each item is represented by a point and a circle. The circle indicates the region of endorsement, that is, a region in the two-dimensional joint plane such that for persons who are positioned inside the circle the probability is larger than a half, whereas for persons positioned outside the circle the probability is smaller than a half. The probability of endorsement becomes smaller the further away the person lies from the circle (or item point). 

In Figure \ref{fig:2d}, the four circles partition the two-dimensional space into 14 regions, where each region corresponds to a predicted response profile, that is, a vector of $R$ zeros and ones indicating which items are endorsed. Thirteen of these regions fall within one or more circles, the fourteenth regions falls outside all circles and corresponds to the profile 0000. Note that the point for this profile can be anywhere outside the circles.
With four items the maximum number of response profiles equals 16, therefore 2 possible response profiles are not represented, which in this case are the profiles where A and D are endorsed without endorsing B and C (1001) and vice versa (0110). More generally, given our proximity model, the maximum number of regions for $R$ items in an $S$-dimensional space is 
\[ 
\genfrac(){0pt}{0}{R-1}{S} + \sum_{s=0}^S \genfrac(){0pt}{0}{R}{s}
\]
\citep{sloane2003line, 1987challenging}. A perfect representation can be found in dimensionality $S = R - 1$. In lower dimensional spaces, not all possible response profiles are represented. Note that in logistic principal component analysis the maximum number of regions for $R$ items in an $S$-dimensional space is \( \sum_{s=0}^S \genfrac(){0pt}{1}{R}{s} \) \citep{coombs1955nonmetric, deleeuw2006principal}, therefore our representation will generally give a better fit.

When predictor variables are available, we constrain the persons points to be functions of these predictors. In this paper, we only consider linear additive functions, that is, $\mathbf{u}_i = \mathbf{B}^\prime\mathbf{x}_i$, but more general nonlinear functions could be considered. In the mapping, these predictor variables will be represented by variable axes with markers. Geometrically, person points can be obtained from these variable axes by the vector sum method also called \emph{interpolation} \citep{gower1996biplots}. 

\subsection{Related Approaches}\label{sec:comparisons}

\cite{desarbo1986simple, desarbo1987constructing} defined the log-odds form as \(\theta_{ir} = m_i - d^2(\mathbf{u}_i, \mathbf{v}_r)\) with \(d^2(\cdot, \cdot)\) the \underline{squared} two-mode Euclidean distance. The use of squared distances changes the geometry and interpretation of the model. Whereas in our mapping the offsets ($m_r$) are related to response variables or items, DeSarbo and Hoffman consider the offsets to be a person characteristic ($m_i$), that is, some persons (observations) have a large \emph{area of acceptance} whereas others have a small area of acceptance. In geometric terms, every person has a circle (with radius $\sqrt{m_i}$) around their position. When an item falls within this circle of a person the probability of a 1 response is higher than 0.5, when it falls outside the circle the probability is smaller than 0.5. Usually, the number of observations $n$ is much larger than the number of response variables or items. Therefore, the number of parameters of the mapping of \cite{desarbo1986simple, desarbo1987constructing} is usually much larger compared to our mapping for the unsupervised case. 

\cite{desarbo1986simple, desarbo1987constructing} also allow for external variables. Like in our approach they constrain $\mathbf{u}_i = \mathbf{B}^\prime\mathbf{x}_i$ in their supervised approach. \cite{desarbo1986simple, desarbo1987constructing} developed and described a conjugate gradient algorithm for estimation of the model parameters. To the best of our knowledge, no software is available anymore to estimate this map. 

\cite{takane1998choice} proposed another closely related model, called MAXSC, where the probabilities are defined as $\pi_{ir} = \exp(\theta_{ir}) / (a + \exp(\theta_{ir})) $ with \(\theta_{ir} = - d^2(\mathbf{u}_i, \mathbf{v}_r)\), also using squared distances. Defining $m^* = \log(a)$ we may write Takane's proposal as
\[
\pi_{ir} = \frac{\exp(- d^2(\mathbf{u}_i, \mathbf{v}_r))}{\exp(m^*) + \exp(- d^2(\mathbf{u}_i, \mathbf{v}_r))}. 
\]
Now dividing numerator and denominator by $\exp(m^*)$ gives
\[
\pi_{ir} = \frac{\exp(- d^2(\mathbf{u}_i, \mathbf{v}_r))/\exp(m^*)}{1 + \exp(- d^2(\mathbf{u}_i, \mathbf{v}_r))/\exp(m^*)} 
\]
and therefore
\[
\pi_{ir} = \frac{\exp(m - d^2(\mathbf{u}_i, \mathbf{v}_r))}{1 + \exp(m - d^2(\mathbf{u}_i, \mathbf{v}_r))}, 
\]
where $m = -m^* = -\log(a)$. Except for the distance or squared distance form, this gives a special case of our mapping with $m_r = m$ for all $r$, but also a special case of the mapping of DeSarbo and Hoffman with $m_i = m$ for all $i$. The threshold parameter $a$ in Takane's model can be considered either a person or item characteristic. In a geometric representation, the threshold can be either included as a one-sized circle around the person points or around the item points. For interpretation, larger values of $a$ lead to smaller probabilities, whereas lower values of $a$ lead to higher probabilities. 

\cite{takane1998choice} did not consider external information, so only investigated the unsupervised case. Whereas in our approach as well as that of DeSarbo and Hoffman the $\mathbf{u}_i$ parameters are fixed effects, Takane assumes them to be random effects with a multivariate normal distribution with mean equal to zero and a diagonal covariance matrix. \cite{takane1998choice} developed an EM-algorithm to maximize the marginal likelihood, with in the M-step a Fisher's scoring algorithm. To the best of our knowledge, no software is available anymore to estimate this map.

A third technique often used for the analysis of proximity items is correspondence analysis \citep{heiser1981unfolding, greenacre1984, terbraak1985correspondence, polak2011, beh2021introduction}. Correspondence Analysis (CA) is an exploratory data analysis tool for tables with non-negative entries. CA decomposes the observed non-negative data as
\[
y_{ir} = y_{++}p_{i+}p_{+r}\left[1 +  \sum_{s = 1}^{S*} f_{is}\lambda_s g_{rs} \right]
\]
subject to the constraints $\sum_i p_{i+}f_{is} = \sum_r p_{+r}g_{rs} = 0$ and $\sum_i p_{i+}f^2_{is} = \sum_r p_{+r}g^2_{rs} = 1$ and where $S*$ is the maximum dimensionality. The next step is to determine an optimal dimensionality $S$, often by an analysis of explained inertia. The fitted values follow the equation above, but the sum over dimensions is from 1 till the optimum $S$. The part belonging to higher dimensions become the residuals. Based on this optimal dimensionality a graphical representation is made of the data. There are several options, we focus on the \emph{row principal normalization}. In this normalization, the row categories are plotted as points with coordinates $u_{is} = \lambda_s f_{is}$, and the column categories as vectors with coordinates $g_{rs}$. In row principal normalization, the Euclidean distances between the row points in the representation approximate differences in centered row profiles, that is, chi-square distances, in the data. The column vectors have a direction and a length. The association with the row categories is reconstructed by projection, and the length indicates how well a column fits the chosen dimensionality. Other normalizations include the column principal normalization, the symmetric normalization where the singular values are evenly distributed over the row and column points, and the principal normalization \citep[see][for details]{greenacre1984}. 

The reason we include CA in our discussion is that \cite{terbraak1985correspondence} showed relationships between correspondence analysis and the following unimodal logistic model
\[
\log\left(\frac{\pi_{ir}}{1 - \pi_{ir}}\right) = m_r -\frac{1}{2}\sum_s (u_{is} - v_{rs})^2/t_r^2 ,  
\]
a model closely related to the squared distance models pointed out above. The $t_r$ parameters are so-called tolerances. 
Ter Braak concludes that under four conditions, that are, equal tolerances ($t_r$), equal or independent maxima ($m_r$), equally spaced or uniformly distributed participants scores ($u_{is}$) and item scores ($v_{rs}$), correspondence analysis provides an approximation to the maximum likelihood solution of this model. In his analysis, the focus of approximation is in terms of the participant ($\mathbf{u}_i$) and item points ($\mathbf{v}_r$). How to exactly translate other estimates of correspondence analysis to, for example, the parameters $m_r$ remains unclear. In Section \ref{sec:applications}, we will show a comparison of correspondence analysis with our model using empirical data. \cite{terbraak1986canonical} extended CA for supervised analysis, by restricting a set of points to be a linear combination of external variables. 


\subsection{Data compression for unsupervised analysis}

Before we delve into the algorithm, we will reorganize the data to improve the efficiency of our algorithm. For the unsupervised case with $R$ response variables only $2^R$ response profiles are possible. For Sugiyama's data about religious practices, for example, $R = 6$ so there are 64 possible response profiles. This provides the opportunity to organize the data into a 64 by 6 matrix instead of a 4,243 by 6 matrix of individual responses. The 64 profiles should be weighted by their frequency of occurrence. We denote this frequency by $n_i$, the number of times the specific profile occurs.  There is one profile that is uninformative, the one with only zeros. In an unsupervised analysis, that profile needs to be removed from the data before the analysis. We will denote by $I$ the number of response profiles, each weighted by $n_i$ for $i = 1, \ldots, I$. The frequencies will be collected in the vector $\mathbf{n}$. Note that the original data can also be cast in this formulation with $n_i = 1$ for all $i$ and $I = n = \sum_i n_i$. 

In the following, we will develop an MM algorithm. In the outer loop of the algorithm, the negative log-likelihood is majorized by a weighted least squares function. In the inner loop, updates of our parameters are computed. 

\subsection{An MM-algorithm for estimating the mapping}\label{sec:algo}

The beauty of MM algorithms is that the idea is quite simple, powerful, and has guaranteed descent. The idea of MM \citep{deleeuw1977convergence, groenen1993majorization, heiser1995convergent, hunter2004tutorial}
for finding a minimum of the function \(\mathcal{L}(\boldsymbol{\theta})\), where \(\boldsymbol{\theta}\) is a
vector of parameters, is to define an auxiliary function, called a
\emph{majorization function},
\(\mathcal{M}(\boldsymbol{\theta}|\boldsymbol{\vartheta})\), with two
characteristics \[
\mathcal{L}(\boldsymbol{\vartheta}) = \mathcal{M}(\boldsymbol{\vartheta}|\boldsymbol{\vartheta})\\
\] 
where \(\boldsymbol{\vartheta}\) is a support point, and 
\[
\mathcal{L}(\boldsymbol{\theta}) \leq \mathcal{M}(\boldsymbol{\theta}|\boldsymbol{\vartheta}).
\] 
The two equations tell us that
\(\mathcal{M}(\boldsymbol{\theta}|\boldsymbol{\vartheta})\) is a
function that lies above (i.e., majorizes) the original function and
touches the original function at the support point. The support point is
defined by the current estimates of $\boldsymbol{\theta}$. The two properties define an iterative sequence for a
convergent algorithm because by construction \[
\mathcal{L}(\boldsymbol{\theta}^+) \leq \mathcal{M}(\boldsymbol{\theta}^+|\boldsymbol{\vartheta}) \leq \mathcal{M}(\boldsymbol{\vartheta}|\boldsymbol{\vartheta}) = \mathcal{L}(\boldsymbol{\vartheta}),
\] where \(\boldsymbol{\theta}^+\) is \[
\boldsymbol{\theta}^+ = \mathrm{argmin}_{\boldsymbol{\theta}} \ \mathcal{M}(\boldsymbol{\theta}|\boldsymbol{\vartheta}),
\] the updated vector or parameters. 

\subsubsection{Outer loop}

Logistic models are often fitted by maximizing the likelihood, or
equivalently minimizing the negative log likelihood, 
\[
\mathcal{L}(\boldsymbol{\theta}) = \sum_{i=1}^I n_i \sum_{r = 1}^R -\log \frac{1}{1 + \exp(-q_{ir}\theta_{ir})} = \sum_{i=1}^I \sum_{r = 1}^R - n_i \log \frac{1}{1 + \exp(-q_{ir}\theta_{ir})} =  \sum_{i=1}^I \sum_{r = 1}^R \mathcal{L}_{ir}(\theta_{ir}),
\] 
with $q_{ir} = 2y_{ir} - 1$ and $\theta_{ir} = m_r - d(\mathbf{u}_i, \mathbf{v}_r)$, as before. Because majorization is closed under summation, we can focus on a single element, \(\mathcal{L}_{ir}(\theta_{ir})\). 

The quadratic majorization theorem states that 
\[
\mathcal{L}_{ir}(\theta_{ir}) \leq \mathcal{L}_{ir}(\vartheta_{ir}) + \mathcal{L}'_{ir}(\vartheta_{ir})(\theta_{ir} - \vartheta_{ir}) + \frac{1}{2}(\theta_{ir} - \vartheta_{ir})H(\theta_{ir} - \vartheta_{ir}),
\] 
for a support point $\vartheta_{ir}$ and for any $H$ that is larger or equal to the second derivative. An upper bound to the second derivative is \(H = \frac{n_i}{4}\).

The first derivative of \(\mathcal{L}_{ir}(\theta_{ir})\) with respect to \(\theta_{ir}\) is 
\[
\begin{aligned}
\mathcal{L}'_{ir}(\theta_{ir}) = \frac{\partial \mathcal{L}_{ir}(\theta_{ir})}{\partial \theta_{ir}} &= - n_i (1 + \exp(-q_{ir}\theta_{ir})) \frac{-q_{ir}\exp(-q_{ir}\theta_{ir})}{-(1 + \exp(-q_{ir}\theta_{ir}))^2} \\ &= - n_i q_{ir} \frac{\exp(-q_{ir}\theta_{ir})}{1 + \exp(-q_{ir}\theta_{ir})} ,
\end{aligned}
\]
such that
\[
\mathcal{L}_{ir}(\theta_{ir}) \leq \mathcal{L}_{ir}(\vartheta_{ir}) - n_iq_{ir} \frac{\exp(-q_{ir}\vartheta_{ir})}{1 + \exp(-q_{ir}\vartheta_{ir})}(\theta_{ir} - \vartheta_{ir}) + \frac{n_i}{8}(\theta_{ir} - \vartheta_{ir})(\theta_{ir} - \vartheta_{ir}).
\]





Define
\(\xi_{ir} = q_{ir} \frac{\exp(-q_{ir}\vartheta_{ir})}{1 + \exp(-q_{ir}\vartheta_{ir})}\),
and work out the majorization function 
\[
\begin{aligned}
\mathcal{L}_{ir}(\theta_{ir}) &\leq \mathcal{L}_{ir}(\vartheta_{ir}) - n_i\xi_{ir}(\theta_{ir} - \vartheta_{ir}) + \frac{n_i}{8}(\theta_{ir} - \vartheta_{ir})(\theta_{ir} - \vartheta_{ir}) \\
&\leq \mathcal{L}_{ir}(\vartheta_{ir}) - n_i\xi_{ir}\theta_{ir} + n_i\xi_{ir}\vartheta_{ir} + \frac{n_i}{8}(\theta_{ir}^2 + \vartheta_{ir}^2 -2\theta_{ir}\vartheta_{ir}) \\
&\leq \mathcal{L}_{ir}(\vartheta_{ir}) + \frac{n_i}{8}\theta_{ir}^2 - n_i\xi_{ir}\theta_{ir} -2\frac{n_i}{8}\theta_{ir}\vartheta_{ir} + n_i\xi_{ir}\vartheta_{ir} + \frac{n_i}{8}\vartheta_{ir}^2 .
\end{aligned}
\] 
Let \(\lambda_{ir} = \vartheta_{ir} + 4\xi_{ir}\) to obtain 
\[
\begin{aligned}
\mathcal{L}_{ir}(\theta_{ir}) &\leq \mathcal{L}_{ir}(\vartheta_{ir}) + \frac{n_i}{8}\theta_{ir}^2  -2\frac{n_i}{8}\theta_{ir}\lambda_{ir}  + \frac{n_i}{8}\lambda_{ir}^2 - \frac{n_i}{8}\lambda_{ir}^2 + n_i\xi_{ir}\vartheta_{ir} + \frac{n_i}{8}\vartheta_{ir}^2 \\
&\leq \mathcal{L}_{ir}(\vartheta_{ir}) + \frac{n_i}{8}(\theta_{ir} - \lambda_{ir})^2 - \frac{n_i}{8}\lambda_{ir}^2 + n_i\xi_{ir}\vartheta_{ir} + \frac{n_i}{8}\vartheta_{ir}^2.
\end{aligned}
\] 
Let us define $ c = \mathcal{L}_{ir}(\vartheta_{ir}) - \frac{n_i}{8}\lambda_{ir}^2 + n_i\xi_{ir}\vartheta_{ir} + \frac{n_i}{8}\vartheta_{ir}^2$, a constant with respect to $\theta_{ir}$, so that we can write
\[
\begin{aligned}
\mathcal{L}_{ir}(\theta_{ir}) &\leq& \frac{1}{8} w_{ir}(\theta_{ir} - \lambda_{ir})^2 + c \\
&\leq& \mathcal{M}_{ir}(\theta_{ir} | \vartheta_{ir}) + c.
\end{aligned}
\] 
with \(w_{ir} = n_i\) for all $r$. 

As \[
\mathcal{L}(\boldsymbol{\theta}) = \sum_{i=1}^I\sum_{r = 1}^R \mathcal{L}_{ir}(\theta_{ir}),
\] 
we have that 
\[
\begin{aligned}
\mathcal{L}(\boldsymbol{\theta}) &\leq \sum_{i=1}^I\sum_{r = 1}^R \frac{1}{8}w_{ir}(\theta_{ir} - \lambda_{ir})^2 + c \\
 & \leq \mathcal{M}(\boldsymbol{\theta}| \boldsymbol{\vartheta}) + c,
\end{aligned}
\] 
so that the majorization function is a weighted least squares function with weights equal to \(w_{ir} = n_i\) for all $r$.

\subsubsection{Inner loop: Minimizing the weighted least squares function}

The geometric structure for \(\theta_{ir}\) equals 
\[
\theta_{ir} = m_r - d(\mathbf{u}_i, \mathbf{v}_r),
\] 
such that the parameters of our optimization function are the offsets $\mathbf{m}$, 
the coordinates $\mathbf{u}_i$ collected in the matrix $\mathbf{U}$ in case of an unsupervised analysis or the regression weights $\mathbf{B}$ for the supervised analyses, and the coordinates $\mathbf{v}_r$ collected in the matrix $\mathbf{V}$. Given current values of these parameters ($\boldsymbol{\vartheta}$), the majorization function becomes 
\[
\mathcal{M}(\mathbf{m}, \mathbf{U}, \mathbf{V}|\boldsymbol{\vartheta}) =  \sum_{i}\sum_r w_{ir}(\lambda_{ir} - m_r + d(\mathbf{u}_i, \mathbf{v}_r))^2.
\]
We will alternate between updating the offsets and the coordinates to find the minimum of our loss function. 

\paragraph{Update of \(m_r\)}\ 

For updating \(m_r\), we consider \(\mathbf{U}\) (or \(\mathbf{B}\)) and \(\mathbf{V}\) as
fixed. Then we need to minimize 
\[
\begin{aligned}
\mathcal{M}(\mathbf{m} |\mathbf{U}, \mathbf{V}, \boldsymbol{\vartheta}) &=  \sum_{i}\sum_r w_{ir}(\lambda_{ir} - m_r + d(\mathbf{u}_i, \mathbf{v}_r))^2 \\
&=  \sum_{i}\sum_r w_{ir}(t_{ir} - m_r)^2, \\
\end{aligned}
\] 
where \(t_{ir} = \lambda_{ir} + d(\mathbf{u}_i, \mathbf{v}_r)\). The
update for \(m_r\) is given by the weighted mean of $t_{ir}$ for every $r$, that is
\[
m_r^+ = \frac{\sum_i w_{ir}t_{ir}}{\sum_i w_{ir}}.
\]

If we would like to constrain $m_1 = m_2 = \ldots, m_R = m$, the update of $m$ becomes
\[
m^+ = \frac{\sum_i \sum_r w_{ir}t_{ir}}{\sum_i \sum_r w_{ir}}.
\]

As a side note, we highlight that person specific offsets ($m_i$), as proposed by \cite{desarbo1986simple, desarbo1987constructing}, could be incorporated in our mapping. Estimates can be obtained as 
\[
m_i^+ = \frac{\sum_r w_{ir}t_{ir}}{\sum_r w_{ir}}.
\]
As noted before, this would increase the number of parameters substantially. It is even possible to go one step further and estimate both item and person specific offsets with even more parameters to estimate. These options are not incorporated in our software.

\paragraph{Update geometric parameters: Unsupervised analysis}\ 

For updating the coordinate matrices we treat \(m_r\) as fixed. 
Let us rewrite our majorization function as 
\begin{align}
\mathcal{M}(\mathbf{U}, \mathbf{V}|\mathbf{m}, \boldsymbol{\vartheta}) &=  \sum_{i} \sum_r w_{ir}(\lambda_{ir} - m_r + d(\mathbf{u}_i, \mathbf{v}_r))^2 \nonumber \\ 
&=  \sum_{i} \sum_r w_{ir}(\delta_{ir} - d(\mathbf{u}_i, \mathbf{v}_r))^2,
\label{eq:uv-function}
\end{align}
where \(\delta_{ir} = -(\lambda_{ir} - m_r)\).

This minimization function in Equation \ref{eq:uv-function} is the usual raw STRESS function often used in multidimensional scaling and unfolding. 
\cite{deleeuw1977applications} and \cite{deleeuw1977convergence} proposed the SMACOF algorithm for minimization of this STRESS function for multidimensional scaling. The SMACOF algorithm is itself an MM algorithm. Convergence properties of this algorithm are described by \cite{deleeuw1988convergence}. 
\cite{heiser1981unfolding, heiser1987joint} showed that multidimensional unfolding can be considered a special case of multidimensional scaling. Subsequently, he developed the SMACOF algorithm to deal with rectangular proximity matrices. Advances in the algorithm are described in \cite{busing2010advances}. An elementary treatment of the algorithm for multidimensional scaling can be found in Chapter 8 of \cite{borg2005modern} and for multidimensional unfolding in Chapter 14. 

In the SMACOF algorithm, the cross product term of the dissimilarities $\delta_{ir}$ with the distances $d(\mathbf{u}_i, \mathbf{v}_r)$ is majorized using the Cauchy-Schwarz inequality by a linear function. \cite{heiser1987joint} defined preliminary updates based on the current $\mathbf{U}$ and $\mathbf{V}$ as
\begin{eqnarray*}
\mathbb{U} &=& \mathbf{P}\mathbf{U}-\mathbf{A}\mathbf{V} \\
\mathbb{V} &=& \mathbf{Q}\mathbf{V}-\mathbf{A}^\prime\mathbf{U}
\end{eqnarray*}
where the matrix $\mathbf{A}$ has elements
\begin{equation}
  a_{ir} = \begin{cases}
    w_{ir} \delta_{ir} / d(\mathbf{u}_i, \mathbf{v}_r), & {\text{if}}\ {d(\mathbf{u}_i, \mathbf{v}_r) > 0},  \\
    0,                                        & {\text{if}}\  {d(\mathbf{u}_i, \mathbf{v}_r) = 0}  
  \end{cases} \label{eq:a-matrix}
\end{equation}
and $\mathbf{P} = \mathrm{diag}(\mathbf{A}\mathbf{1})$, and $\mathbf{Q} = \mathrm{diag}(\mathbf{1}^\prime\mathbf{A})$. Collecting the weights $w_{ir}$ in the matrix $\mathbf{W}$ and defining the diagonal matrices $\mathbf{R} = \mathrm{diag}(\mathbf{W}\mathbf{1})$ and $\mathbf{C} = \mathrm{diag}(\mathbf{1}^\prime\mathbf{W})$, the majorization function for the raw STRESS function (Equation \ref{eq:uv-function}) is given by \citep{heiser1987joint, busing2010advances}
\begin{equation}
\mathcal{M}(\mathbf{U}, \mathbf{V}|\mathbf{m}, \boldsymbol{\vartheta}) \leq  \sigma^2(\mathbf{U},\mathbf{V}) = c + \mathrm{tr}\left( \mathbf{U}^\prime\mathbf{R}\mathbf{U} \right) + \mathrm{tr}\left( \mathbf{V}^\prime\mathbf{C}\mathbf{V} \right) - 2 \mathrm{tr}\left( \mathbf{U}^\prime\mathbf{W}\mathbf{V} \right) 
    - 2 \mathrm{tr} \left( \mathbf{U}^\prime \mathbb{U} \right) 
    - 2 \mathrm{tr} \left( \mathbf{V}^\prime \mathbb{V} \right). \label{eq:twomode}
\end{equation}
Alternating between updates for $\mathbf{U}$ and $\mathbf{V}$ provides the minimum for (\ref{eq:twomode}):
keeping $\mathbf{V}$ fixed, the update for $\mathbf{U}$ is given as
\[
\mathbf{U} = \mathbf{R}^{-1} \left( \mathbb{U} + \mathbf{W}\mathbf{V} \right) 
\] 
and keeping $\mathbf{U}$ fixed, the update for $\mathbf{V}$ is given as
\[
\mathbf{V} = \mathbf{C}^{-1} \left( \mathbb{V} + \mathbf{W}^\prime\mathbf{U}\right).
\]

This standard SMACOF algorithm, as just described, was derived under the assumption that the (working) dissimilarities are non-negative. However, in our case we cannot guarantee that this assumption is true in every cycle of the algorithm. 
\cite{heiser1991generalized} showed a way to deal with negative dissimilarities in multidimensional scaling.
We will generalize that approach to the two-mode distance case. 

The line of thought of Heiser's contribution is that two majorizing functions are defined: one for the case that the dissimilarity is non-negative and one for the case that the dissimilarity is negative. When the dissimilarities are non-negative, we can majorize, as described above, by a linear function. When the dissimilarities are negative, the STRESS function can be majorized by a quadratic function. \cite{heiser1991generalized} showed that the updating formulae are still valid but that some elements of the matrices $\mathbf{W}$ and $\mathbf{A}$ need to be defined differently, depending on the sign of the dissimilarity.
Matrix $\mathbf{W} = \{ w_{ir} \}$ is redefined as
\begin{eqnarray*}
  w_{ir} = \left\{ \begin{array}{ll} 
    w_{ir}                                                                                              & \mathrm{if}\ \delta_{ir} \geq 0,\\
    \left[ w_{ir}(d(\mathbf{u}_i, \mathbf{v}_r) + |\delta_{ir}|)\right] / d(\mathbf{u}_i, \mathbf{v}_r) & \mathrm{if}\ \delta_{ir} < 0\  \mathrm{and}\ d(\mathbf{u}_i, \mathbf{v}_r) > 0, \\
    \left[ w_{ir}(\epsilon + \delta_{ir}^2) \right] / \epsilon                                          & \mathrm{if}\ \delta_{ir} < 0\  \mathrm{and}\ d(\mathbf{u}_i, \mathbf{v}_r) = 0,
  \end{array}\right. 
\end{eqnarray*} 
where \(\epsilon\) is a small positive constant. The elements of $\mathbf{A}$ depend on the sign of the working dissimilarities, that is, for negative $\delta_{ir}$ the corresponding $a_{ir}$ are set to zero, while for non-negative $\delta_{ir}$ the $a_{ir}$ are defined as in Equation \ref{eq:a-matrix}.

Within the inner loop we could iterate till convergence, but a few iterations of updating $\mathbf{U}$ and $\mathbf{V}$ are enough. In other words, instead of finding the minimum of the majorization function in every iteration we only need to take a few steps in the right direction. This iterative sequence still guarantees convergence to a (local) minimum.

Note that, the unsupervised mapping has rotational and translational freedom. 
To obtain identified solutions, we require $\sum_i n_i u_{is} = 0$ for every dimension $s$ to curb the translational freedom. 
To deal with the rotational freedom, we rotate the solution such that $\mathbf{U}$ is in principal coordinates. Therefore, we define the diagonal matrix $\mathbf{D}_n$ with elements $d_{ii} = n_i$ and perform an eigenvalue decomposition $\mathbf{U}^\prime\mathbf{D}_n \mathbf{U} = \mathbf{E}\bs{\Phi}\mathbf{E}^\prime$ and rotate (i.e., post multiply) both $\mathbf{U}$ and  $\mathbf{V}$ by $\mathbf{E}$.

\paragraph{Update geometric parameters: Supervised analysis}\ 

\cite{deleeuwheiser1980} considered multidimensional scaling with restrictions on the configuration, like we use in the supervised mapping. \cite{heiser1987joint} described how to use linear constraints in multidimensional unfolding, which was further developed in \cite{busing2010restricted}. 

In the supervised analysis, we have that $\mathbf{U} = \mathbf{XB}$ and we need to estimate $\mathbf{B}$ instead of $\mathbf{U}$. Given the other parameters, the update of $\mathbf{B}$ is given as 
\[
\mathbf{B} = \left(\mathbf{X}^\prime\mathbf{R}\mathbf{X} \right)^{-1} \left(\mathbf{X}^\prime \mathbb{U} + \mathbf{X}^\prime\mathbf{W}\mathbf{V}\right)
\] 
and before updating $\mathbf{V}$ we compute $\mathbf{U} = \mathbf{XB}$.

For the supervised mapping there is only rotational freedom because the origin corresponds to $\mathbf{x} = \mathbf{0}$. Like in the unsupervised analysis we rotate the solution such that $\mathbf{U}$ is in principal coordinates. We find the rotation matrix $\mathbf{E}$ as in the unsupervised case and rotate both $\mathbf{B}$ and $\mathbf{V}$ by $\mathbf{E}$. 

\subsection{Local Optima and Initialisation}

We need to remark that the loss function is consistently minimized by this algorithm. However, the algorithm does not guarantee that the obtained minimum is the global minimum. This so-called convergence to local minimum problem may be mitigated by either choosing good (or rational) initial parameter values or by using many random starts. Good rational starting values can be obtained by performing, for example, a (canonical) correspondence analysis. However, even these good starts do not guarantee to find the global minimum. Therefore, it is recommended to also perform a number of random starts. 
In our implementation, we draw elements of the parameter matrices $\mathbf{U}$ or $\mathbf{B}$ and $\mathbf{V}$ from independent standard normal distributions as initial values. For the offset parameter, we use the average of the $q_{ir}$ as initial values for $m_r$. 

\subsection{Algorithm Scheme and Implementation}

An overview of the algorithms for the unsupervised and supervised mappings can be found in Algorithm \ref{alg:lmdu} and Algorithm \ref{alg:lrmdu}, respectively. Note that the matrix $\bs{\Pi}$ has elements $\pi_{ir}$. We implemented the algorithm in the R-package \texttt{lmap} \citep{lmappackage}. For the estimation of the map, the function \texttt{lmdu} can be used. The package also has a function for plotting the map. 

\begin{minipage}{0.46\textwidth}
\begin{algorithm}[H]
  \tiny
  \caption{Unsupervised Algorithm}\label{alg:lmdu}
  \begin{algorithmic}[1]
  \State Input: $\mathbf{Y}, \mathbf{n}, \mathbf{m}, \mathbf{U}, \mathbf{V}, S$
  \State predefine: maxouter, maxinner, $\epsilon_1$, $\epsilon_2$
  \State assess $\mathcal{L}^0(\mathbf{m}, \mathbf{U}, \mathbf{V})$                       
  \For {$t_1 \leftarrow 1,\text{maxouter}$}                                                                                    
    \State compute $\mathbf{\Pi}$                                            
    \State compute $\bs{\Lambda} \leftarrow \mathbf{1m}^\prime - d(\mathbf{U},\mathbf{V}) + 4 (\mathbf{Y} - \bs{\Pi})$
    \State compute $\mathbf{T} \leftarrow \bs{\Lambda} + d(\mathbf{U}, \mathbf{V})$
    \State compute $\mathbf{m} \leftarrow (\mathbf{W} \odot \mathbf{T})^\prime \mathbf{1}/n$                    
    \State assess $\mathcal{M}^{0}(\mathbf{m}, \mathbf{U}, \mathbf{V}|\bs{\vartheta})$                 
    \For {$t_2 \leftarrow 1,\text{maxinner}$}                                                                              
      \State compute $\mathbf{A}$ and $\mathbf{P}, \mathbf{Q}$               
      \State compute $\mathbf{W}$ and $\mathbf{R}, \mathbf{C}$               
      \State compute $\mathbf{U} \leftarrow \mathbf{R}^{-1} \left(\mathbf{P}\mathbf{U} - \mathbf{A}\mathbf{V} + \mathbf{W}\mathbf{V}\right)$
      \State compute $\mathbf{V} \leftarrow \mathbf{C}^{-1} \left(\mathbf{Q}\mathbf{V} - \mathbf{A}^\prime\mathbf{U} + \mathbf{W}^\prime\mathbf{U} \right)$
      \State assess $\mathcal{M}^{t_2}(\mathbf{m}, \mathbf{U}, \mathbf{V}|\bs{\vartheta})$                 
      \State if $\mathcal{M}^{t_2}(\mathbf{m}, \mathbf{U}, \mathbf{V}|\bs{\vartheta}) - \mathcal{M}^{t_2-1}(\mathbf{m}, \mathbf{U}, \mathbf{V}|\bs{\vartheta}) < \epsilon_1$: break
    \EndFor
    \State assess $\mathcal{L}^{t_1}(\mathbf{m}, \mathbf{U}, \mathbf{V})$                      
    \State if $\mathcal{L}^{t_1}(\mathbf{m}, \mathbf{U}, \mathbf{V}) - \mathcal{L}^{t_1-1}(\mathbf{m}, \mathbf{U}, \mathbf{V}) < \epsilon_2$: break
  \EndFor
  \State eigenvalue decomposition $\mathbf{U}^\prime\mathbf{D}_n\mathbf{U}$: $\mathbf{E}\mathbf{\Phi}\mathbf{E}^\prime$
  \State rotate $\mathbf{U} \leftarrow \mathbf{U}\mathbf{E}$ and $\mathbf{V} \leftarrow \mathbf{V}\mathbf{E}$
  \State return($\mathbf{m}, \mathbf{U},\mathbf{V}$)
  \end{algorithmic}
\end{algorithm}
\end{minipage}
\hfill
\begin{minipage}{0.46\textwidth}
\begin{algorithm}[H]
  \tiny
  \caption{Supervised Algorithm}\label{alg:lrmdu}
  \begin{algorithmic}[1]
  \State Input: $\mathbf{Y}, \mathbf{X}, \mathbf{n}, \mathbf{m}, \mathbf{B}, \mathbf{V}, S$
  \State predefine: maxouter, maxinner, $\epsilon_1$, $\epsilon_2$
  \State assess $\mathcal{L}^0(\mathbf{m}, \mathbf{B}, \mathbf{V})$                       
  \For {$t_1 \leftarrow 1,\text{maxouter}$}                                                                                    
    \State compute $\mathbf{\Pi}$                                            
    \State compute $\bs{\Lambda} \leftarrow \mathbf{1m}^\prime - d(\mathbf{XB},\mathbf{V}) + 4 (\mathbf{Y} - \bs{\Pi})$
    \State compute $\mathbf{T} \leftarrow \bs{\Lambda} + d(\mathbf{XB}, \mathbf{V})$
    \State compute $\mathbf{m} \leftarrow (\mathbf{W} \odot \mathbf{T})^\prime \mathbf{1}/n$                    
    \State assess $\mathcal{M}^{0}(\mathbf{m}, \mathbf{XB}, \mathbf{V}|\bs{\vartheta})$                 
    \For {$t_2 \leftarrow 1,\text{maxinner}$}                                                                              
      \State compute $\mathbf{A}$ and $\mathbf{P}, \mathbf{Q}$               
      \State compute $\mathbf{W}$ and $\mathbf{R}, \mathbf{C}$               
      \State compute $\mathbf{B} \leftarrow \left(\mathbf{X}^\prime\mathbf{R}\mathbf{X} \right)^{-1} \left[\mathbf{X}^\prime \left(\mathbf{P}\mathbf{XB}- \mathbf{A}\mathbf{V}\right) + \mathbf{X}^\prime\mathbf{W}\mathbf{V}\right]$
      \State compute $\mathbf{V} \leftarrow \mathbf{C}^{-1} \left(\mathbf{Q}\mathbf{V} - \mathbf{A}^\prime\mathbf{XB} + \mathbf{W}^\prime\mathbf{XB} \right)$
      \State assess $\mathcal{M}^{t_2}(\mathbf{m}, \mathbf{XB}, \mathbf{V}|\bs{\vartheta})$                 
      \State if $\mathcal{M}^{t_2}(\mathbf{m}, \mathbf{XB}, \mathbf{V}|\bs{\vartheta}) - \mathcal{M}^{t_2-1}(\mathbf{m}, \mathbf{XB}, \mathbf{V}|\bs{\vartheta}) < \epsilon_1$: break
    \EndFor
  \State assess $\mathcal{L}^{t_1}(\mathbf{m}, \mathbf{B}, \mathbf{V})$                      
  \State if $\mathcal{L}^{t_1}(\mathbf{m}, \mathbf{B}, \mathbf{V}) - \mathcal{L}^{t_1-1}(\mathbf{m}, \mathbf{B}, \mathbf{V}) < \epsilon_2$: break
  \EndFor
  \State eigenvalue decomposition $\mathbf{B}^\prime\mathbf{X}^\prime\mathbf{D}_n\mathbf{XB}$: $\mathbf{E}\mathbf{\Phi}\mathbf{E}^\prime$
  \State rotate $\mathbf{B} \leftarrow \mathbf{B}\mathbf{E}$ and $\mathbf{V} \leftarrow \mathbf{V}\mathbf{E}$
  \State return($\mathbf{m}, \mathbf{B},\mathbf{V}$)
  \end{algorithmic}
\end{algorithm}
\end{minipage}

\subsection{Model Selection and Assessment}

For the application of our mapping (both the unsupervised as well as the supervised) to empirical data, the user has to define or choose a dimensionality $S$. Although we have a likelihood method, likelihood ratio statistics cannot be used for dimensionality selection \citep{takane2003likelihood}, because a regularity condition for this statistic to be chi-square distributed is not satisfied. To find an optimal dimensionality, we will use the AIC. AIC is based on the entropic or information-theoretic interpretation of the maximum likelihood method as well as the minimization of the Kullback-Leibler information quantity \citep{akaike1974new, burnham2004multimodel, anderson2007model}. The AIC for any model can be defined as
\[
    \mathrm{AIC} =  2 \hat{\mathcal{L}}(\bm{\theta}) + 2\mathrm{npar},
\]
where $\mathrm{npar}$ denotes the number of parameters of the model. The first term $2\hat{\mathcal{L}}(\bm{\theta})$ in AIC is twice the negative log likelihood (usually called the deviance) and it acts as a measure of lack of fit to the data, for which smaller values will be preferred. The second term, $2\mathrm{npar}$, acts as a penalty term which penalizes complex models for having many parameters. The aim is to reach a balance between the lack of fit and the model complexity: models with smaller AIC values will indicate a better balance. The optimal model choice minimizes the AIC. For the AIC, the number of parameters is needed. For the unsupervised case the number of parameters is 
$\mathrm{npar} = (I-1)S + RS + R - S(S-1)/2$, while for supervised analysis it is $\mathrm{npar} = PS + RS + R - S(S-1)/2$.  

For supervised models, there is also the question which predictor variables have an effect on the response variables. Again there is a problem with the likelihood ratio statistic. It compares the fit of two nested models by dividing the likelihood value obtained under a null hypothesis by the likelihood obtained under an alternative hypothesis. If the model under the null hypothesis is true and certain regularity conditions are satisfied, minus two times the log of this ratio is known to be asymptotically distributed as a chi-square variable with degrees of freedom equal to the difference in the number of parameters under the two hypotheses. The assumption that the model under the null hypothesis should be true is problematic in our case. Suppose that we would like to test whether the first predictor has an effect on the responses and therefore estimate the model with and without this first predictor. Even if the first predictor has no effect on the outcomes, the model can be miss-specified in many other aspects. One example is that some of the other predictors have a nonlinear effect on the log odds of some of the response variables. Therefore, we also propose to use the AIC for selection of predictor variables. 

For the assessment of model fit, we can evaluate several statistics. First, we can look at different types of residuals. The simplest type of residuals are the \emph{raw residuals}, $e_{ir} = y_{ir} - \hat{\pi}_{ir}$. These residuals are positive for participants with $y_{ir} = 1$ and negative for the others. The range of these raw residuals is from -1 to 1. 

Another type of residuals are the \emph{deviance residuals}. These deviance residuals partition the total deviance in small pieces, such that the sum of squared deviance residuals equal the total deviance. These deviance residuals are defined as
$$
\mathcal{D}_{ir} = \mathrm{sign}(e_{ir}) \sqrt{2 \mathcal{L}_{ir}}.
$$
These residuals are a by-product of the estimation procedure, as the $\mathcal{L}_{ir}$ are computed in every iteration. From the deviance residual, we can evaluate the contribution of each participants or each item to the overall deviance, by squaring and summing over items or participants, respectively. 

In the supervised mapping, we make an assumption about the shape of the relationship between the predictor variables and the response variable. Whereas in linear models the implied functional form is linear, in our mapping the functional form is defined by the distance. To verify whether this functional form is correct, we will generalize component plus residual plots. Such plots have been proposed for linear and logistic regression, see \cite{fox2015applied}. For our mapping, such a plot is created for each combination of a predictor and a response variable. On the horizontal axis the predictor variable is depicted. On the vertical axis we show the partial fit plus four times the raw residual ($e_{ir}$). The partial fit is defined as $m_r - d(\bm{x}_p\bm{b}'_p, \bm{v}_r)$. We add four times the raw residual, as this is also used in our algorithm (as shown in Section \ref{sec:algo}).  We add two lines to this scatterplot: the first line shows the assumed functional form, the second is a  smooth loess curve \citep{cleveland1979robust}. When the two lines strongly deviate from each other this indicates model misspecification. 

One aspect of model assessment is to check for influential cases. To detect the influence of observations on the fit, usually the model is fitted $n$ times, every time leaving one participant out. Indicators of influence are given by the (standardized) change in estimated parameters or fit measures. For linear regression models it has been shown that such statistics can be derived without actually having to refit the model $n$ times. For logistic regression, \cite{pregibon1981logistic} derived approximations to these indicators. As computational power has increased dramatically over the last decades, we will not develop approximations, but use the computational power to find indicators of influence. The model will be estimated by leaving out participant $i$. The estimated values when leaving out an observation will be denoted by $\hat{\mathbf{B}}_{-i}$, and similarly for other parameters. We propose to use the following indicators. First the change in deviance
$$
\delta_{\mathcal{D}}(i) = 2\left( \mathcal{L}(\bm{m}, \bm{B}, \bm{V}) - \mathcal{L}(\bm{m}_{-i}, \bm{B}_{-i}, \bm{V}_{-i}) \right)
$$
which is similar to Cooks distance \citep{williams1987generalized}. Note that the likelihood is evaluated on the complete sample using the estimates obtained when leaving participant $i$ out. Second, the overall change in regression weights is defined by
$$
\delta_B(i) = \| \hat{\mathbf{B}} - \hat{\mathbf{B}}_{-i} \|^2,
$$
where $\| \bm{Q} \|^2$ takes the sum of squares of all elements of the matrix $\bm{Q}$. 
Finally, the overall change in item locations is defined by the following measure
$$
\delta_V(i) = \| \hat{\mathbf{V}} - \hat{\mathbf{V}}_{-i} \|^2.
$$
For all three indicators, higher values represent larger influence of observation $i$. 

\section{Applications}\label{sec:applications}

In this section, we show applications on two empirical data sets. The first data set is Sugiyama's data on religious practices, the second data set concerns vote intentions for parliamentary elections in The Netherlands.

We apply our unsupervised mapping to the first data set and compare the results against correspondence analysis. For the vote intention data, we first apply an unsupervised mapping and thereafter include the predictor variables in the supervised mapping. Before we delve into the applications, we discuss for all three analyses the occurrence of local optima. Afterwards, we interpret the solutions for both data sets.

\subsection{Severity of Local Optima}

\begin{figure}[t]
\noindent\begin{subfigure}[b]{0.33\textwidth}
\includegraphics[width = .9\textwidth]{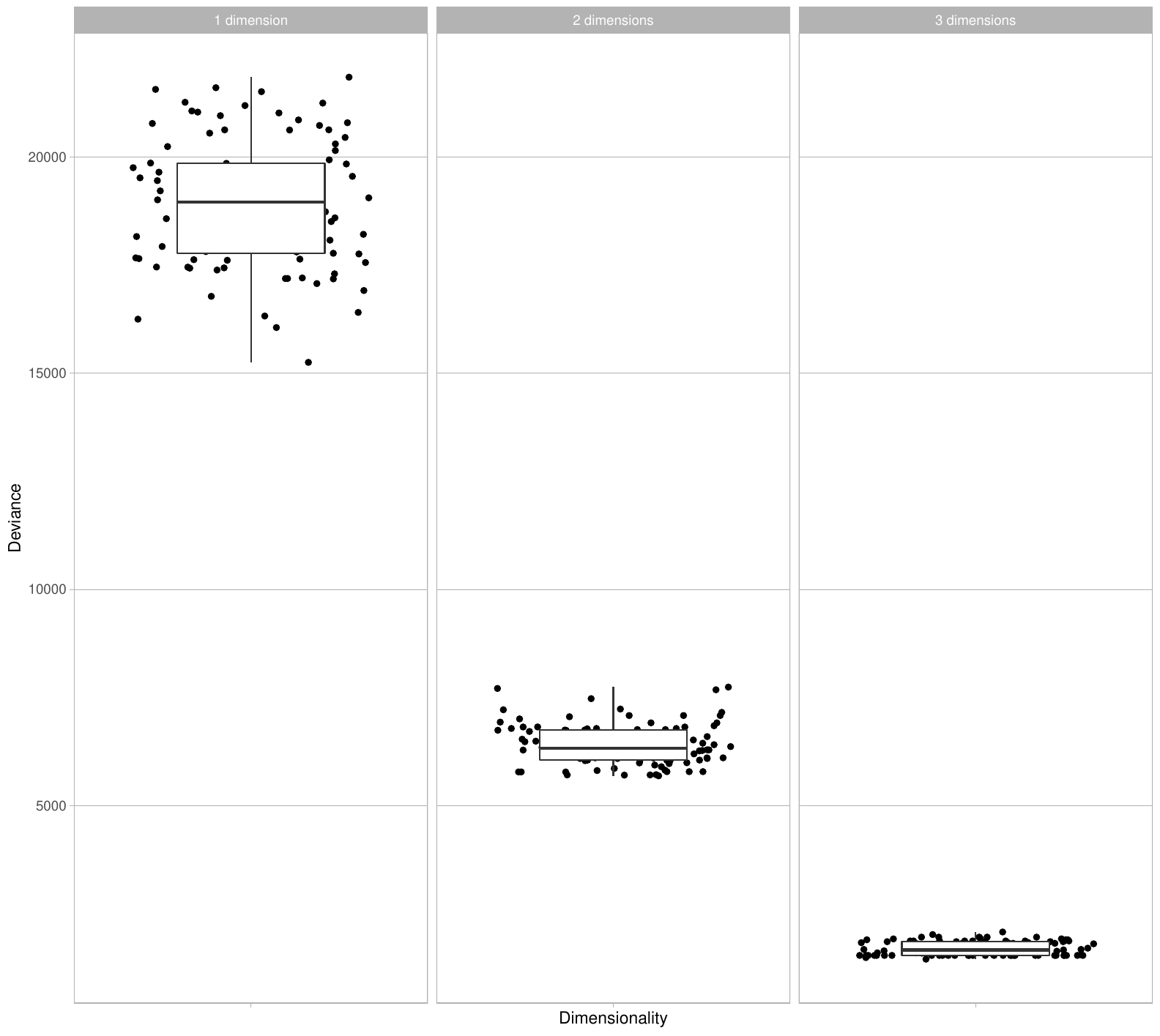}
\caption{}
\label{fig:locoptsugi}
\end{subfigure}%
\noindent\begin{subfigure}[b]{0.33\textwidth}
\includegraphics[width = .9\textwidth]{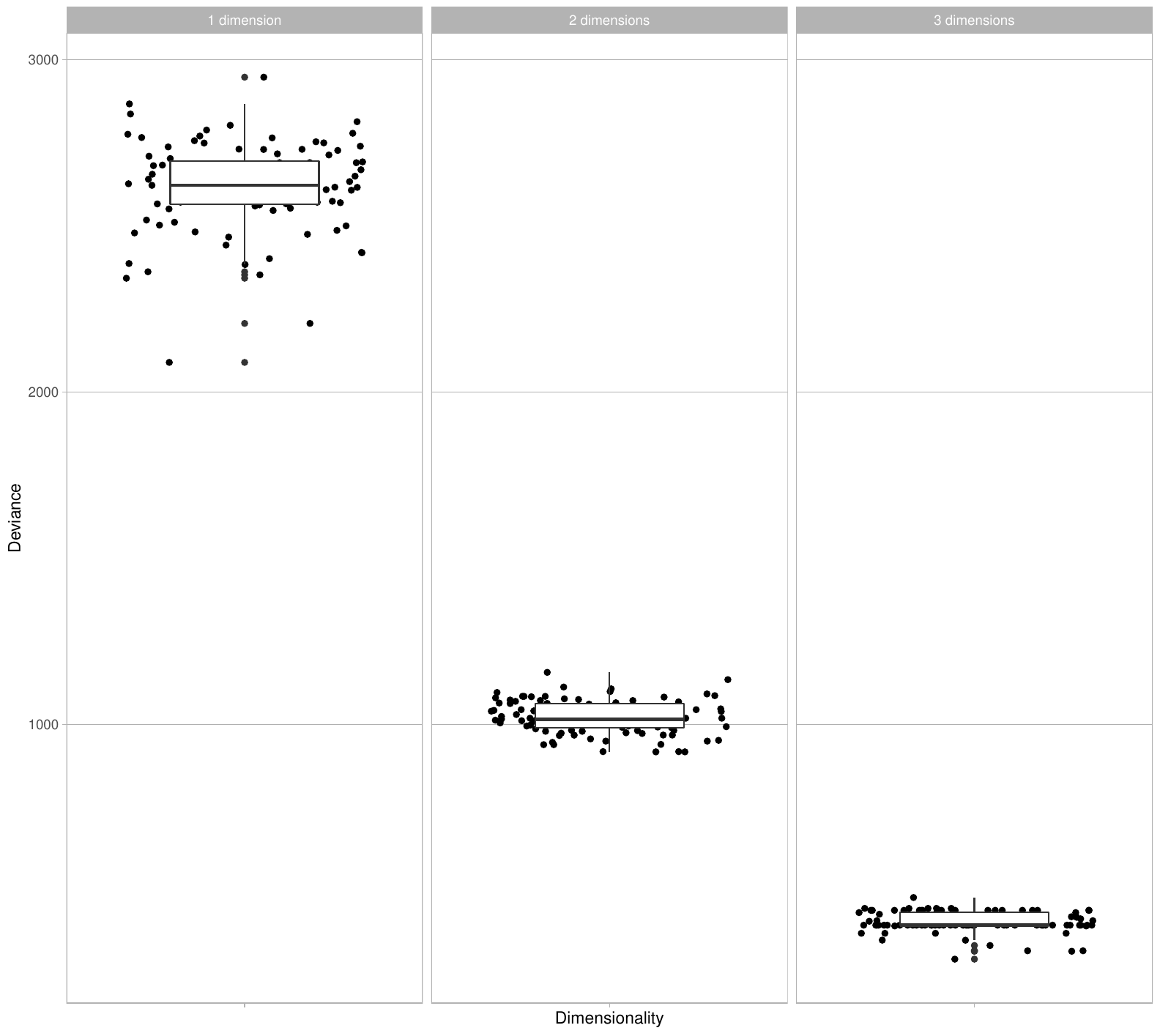}
\caption{}
\label{fig:locoptdpes1}
\end{subfigure}%
\noindent\begin{subfigure}[b]{0.33\textwidth}
\includegraphics[width = .9\textwidth]{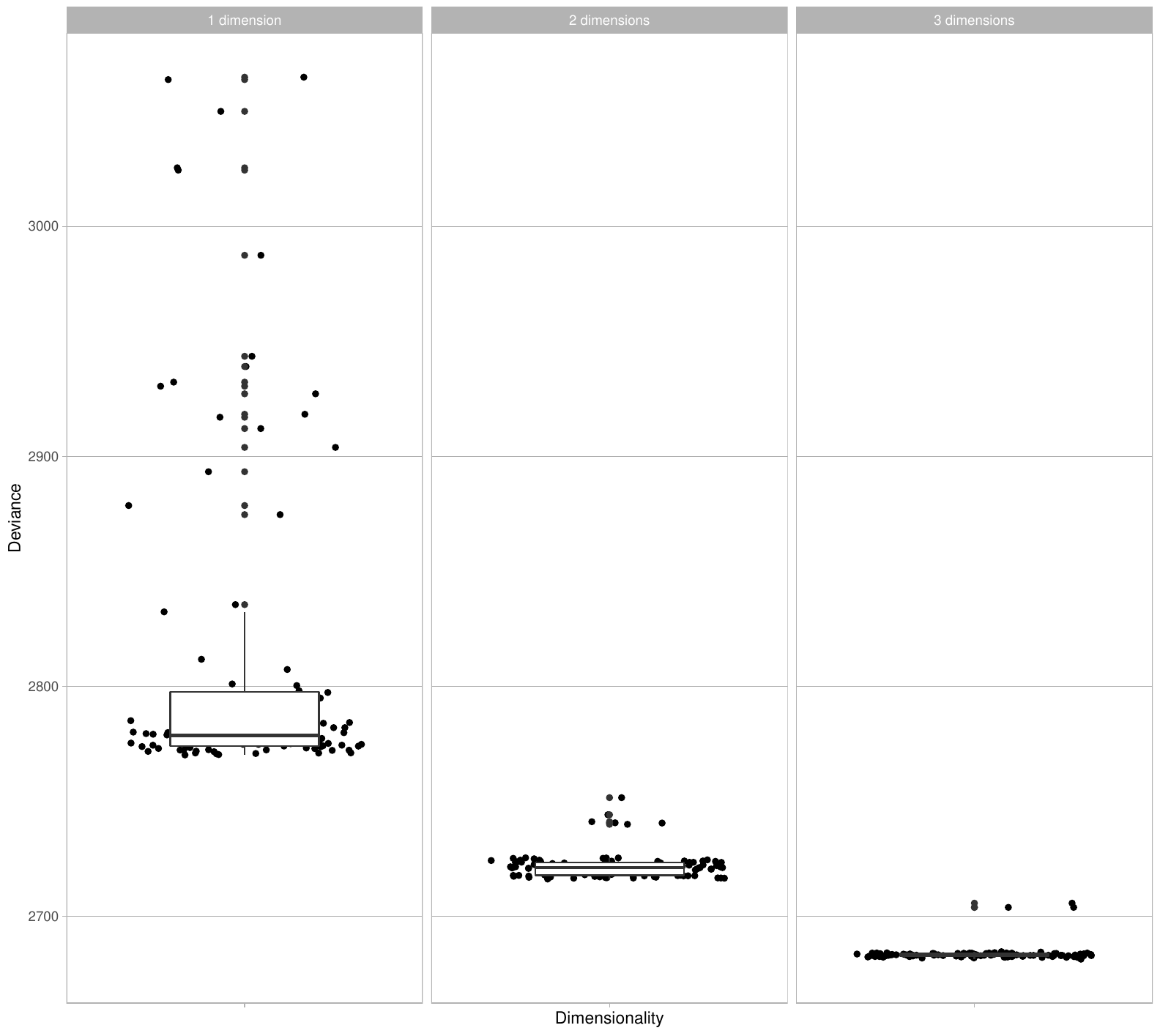}
\caption{}
\label{fig:locoptdpes2}
\end{subfigure}%
\caption{Deviances of 100 random starts in 1, 2, and 3 dimensions for 
(a) Sugiyama's data, 
(b) the unsupervised analysis of the Dutch election data, and 
(c) the supervised analysis of the Dutch election data. A little bit of horizontal jitter is used to better visualize the deviance of the 100 solutions.}
\end{figure}

For each of the three analyses, we performed 100 different starts in dimensionalities 1, 2, and 3. The deviances (twice the negative loglikelihood) of these analyses are shown in Figure 3, for the unsupervised analysis of Sugiyama's data in Figure \ref{fig:locoptsugi}, for the unsupervised analysis of the Dutch election data in Figure \ref{fig:locoptdpes1}. and for the supervised analysis of the Dutch election data in 
Figure \ref{fig:locoptdpes2}. Generally, we see that local optima occur in all dimensionalities, but that especially the unidimensional analyses are prone to local optima. The local optima problem for the supervised analysis seems to be less severe as the median is very close to the minimum.



\subsection{Sugiyama's data on Religious Practices}

For Sugiyama's data on religious practices, the respondents had to answer \emph{yes} or \emph{no} to the following 6 questions:
\begin{itemize}
\item[A] Do you make it a rule to practice religious conduct, such as attending religious services, religious worship, and missionary works and do you occasionally offer prayers or chant sutras?
\item[B] Do you visit a grave once or twice a year?
\item[C] Do you occasionally read religious books, such as the Bible or the Buddhist Scriptures?
\item[D] Do you visit shrines and temples to pray for business prosperity, success in an entrance examination, and so forth?
\item[E] Do you keep a talisman, such as an amulet, charm, or mascot near you?
\item[F] Did you draw a fortune, consult a diviner, or had you your fortune told within the last year?
\end{itemize}
The 64 answer patterns with response frequencies can be found in \cite{heiser1981unfolding} and \cite{takane1998choice}. Three response patterns do not occur in the data. In total, there were 4243 participants. Because all zeros are uninformative, the pattern with only zeros is left out of the analysis. This reduces the sample size with 718 to 3525.

The deviance of the intercepts only model (i.e., $\theta_{ir} = m_r$) equals 24,011.13

The smallest deviance in one dimension is 15,254.3, in two dimensions 5,690.7, and in three dimensions 1,451.8, so that 36.5\%, 76.3\% and 93.9\% of the deviance is explained by these three mappings, respectively. The AICs are 22328.3, 19824.7, and 22643.8 for the one, two, and three dimensional solution, respectively. 
The maximum number of represented profiles are 12, 32, and 52 for the uni-, two-, and three-dimensional solution, respectively. We will further look at the two-dimensional solution.

\begin{figure}
    \centering
    \begin{minipage}{.5\textwidth}
        \centering
       \includegraphics[width = 1.15\textwidth]{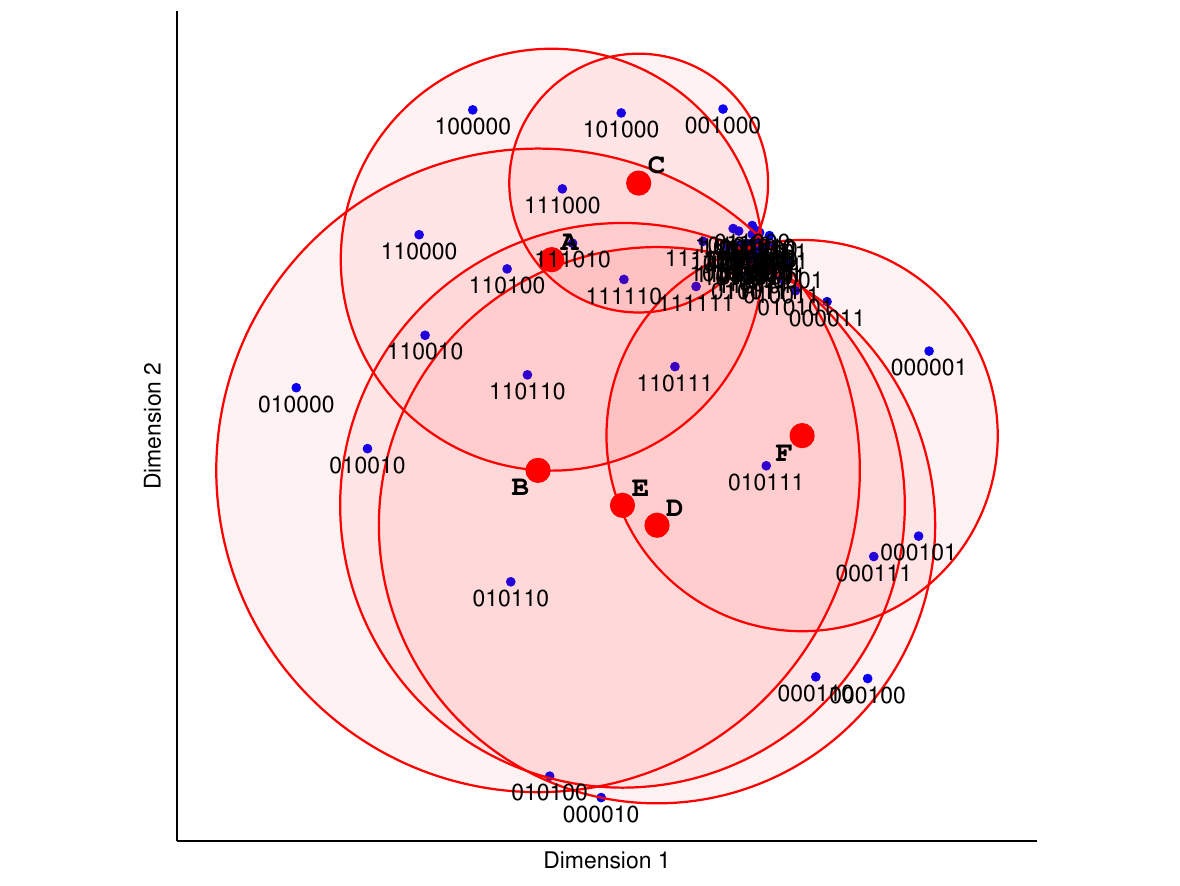}
    \end{minipage}%
    \begin{minipage}{0.5\textwidth}
        \centering
        \includegraphics[width = 1.15\textwidth]{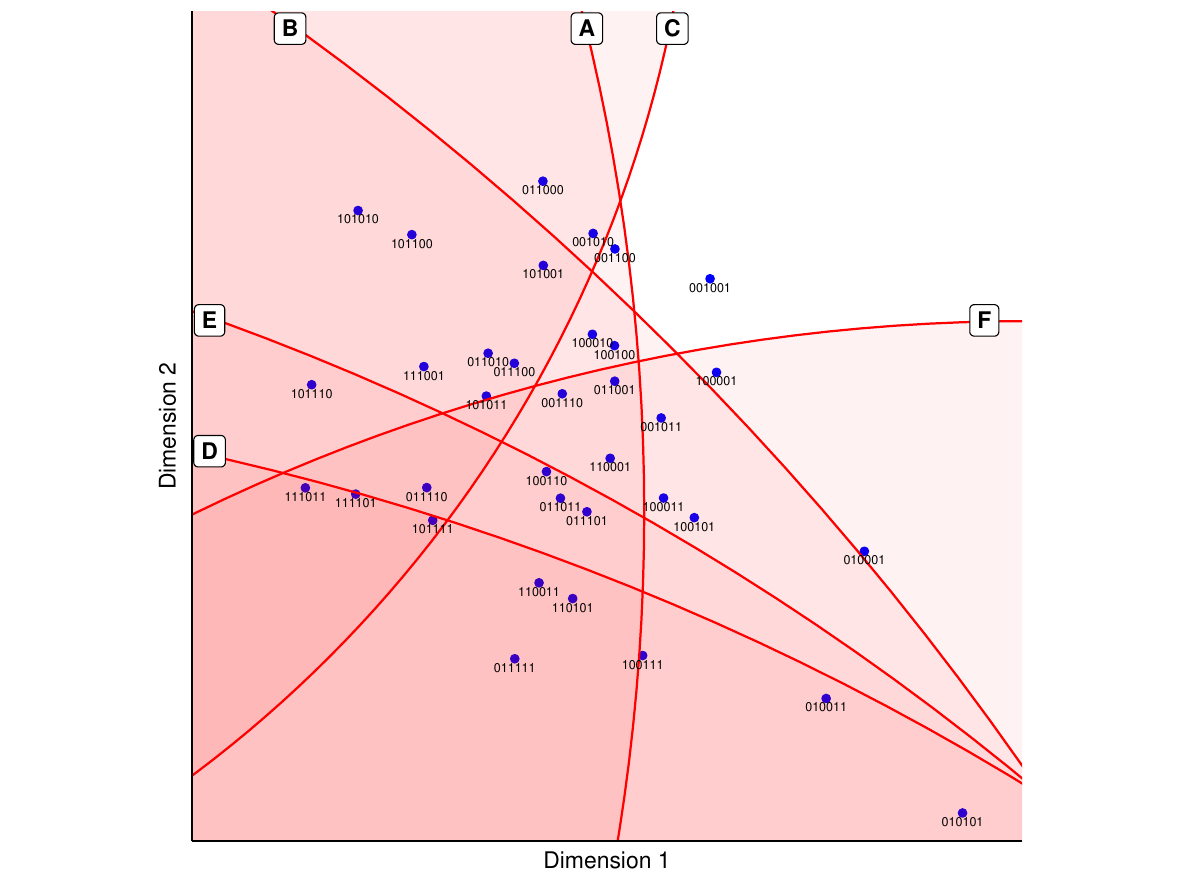}
    \end{minipage}
    \caption{Two-dimensional solution for Sugiyama's data. The six religious practice items are indicated by the letters A till F. Subject points are labelled with their response patterns. Left panel: the estimated configuration; right panel: a zoom into the clutter of response profiles.}
    \label{fig:sugi}
\end{figure}

The two dimensional solution is displayed in the left panel of Figure \ref{fig:sugi}, where each of the six items is represented by a point with coordinate $\mathbf{v}_r$ and a circle with radius $m_r$. Persons with locations inside the circle have a probability larger than 0.5 for the corresponding item. We see that item C has the smallest circle, that is having a relatively small  area of endorsement, while item B has the largest circle. We shaded the regions of endorsement in such a way that when two (or more) circles overlap the shading becomes darker. 

Considering the positions of the items, we see that items B, E, and D lie close together, as well as A and C, while item F is a bit isolated. We see that the circles for items C and F only slightly overlap, indicating that these two items are seldom picked together. Circles for items E and D, however, do overlap considerably indicating that these two items are often picked together. The person points are labeled by the response pattern. For example, the point labelled as 110010 represent participants that picked items A, B, and E while not picking items B, C, and F. We see that the response patterns with a single 1 fall on the outside of the joint space, within a single circle. For example, at the top there are the response patterns for only A (100000) and only C (001000), the points fall within the regions of endorsement of the corresponding items. In the middle of these two points is the response pattern for both A and C (101000), falling in the region of endorsement of both items. In this representation 23 profiles can be seen clearly, where many other profiles clutter together. These 23 profiles represent 2811 participants, that is 79.7\% of the respondents. These 23 profiles all fall in the correct regions. 

There is a large bulk of subject points near the intersection of all circles, representing mainly answer profiles with 2, 3, and 4 picks.  In the right hand panel of Figure \ref{fig:sugi}, we zoom into this region with many response patterns. Note that many regions, contain multiple profiles points, such that relatively many of these profiles are missclassified. These missclassified points, however, only represent a small part of the data (714 participants) and response profiles with small weights. The profile with most participants in this zoomed-in visualization is 010001 (on the right hand side of the figure) and has 137 participants.

%

\begin{figure}
    \centering
        \includegraphics[width=1.1\textwidth]{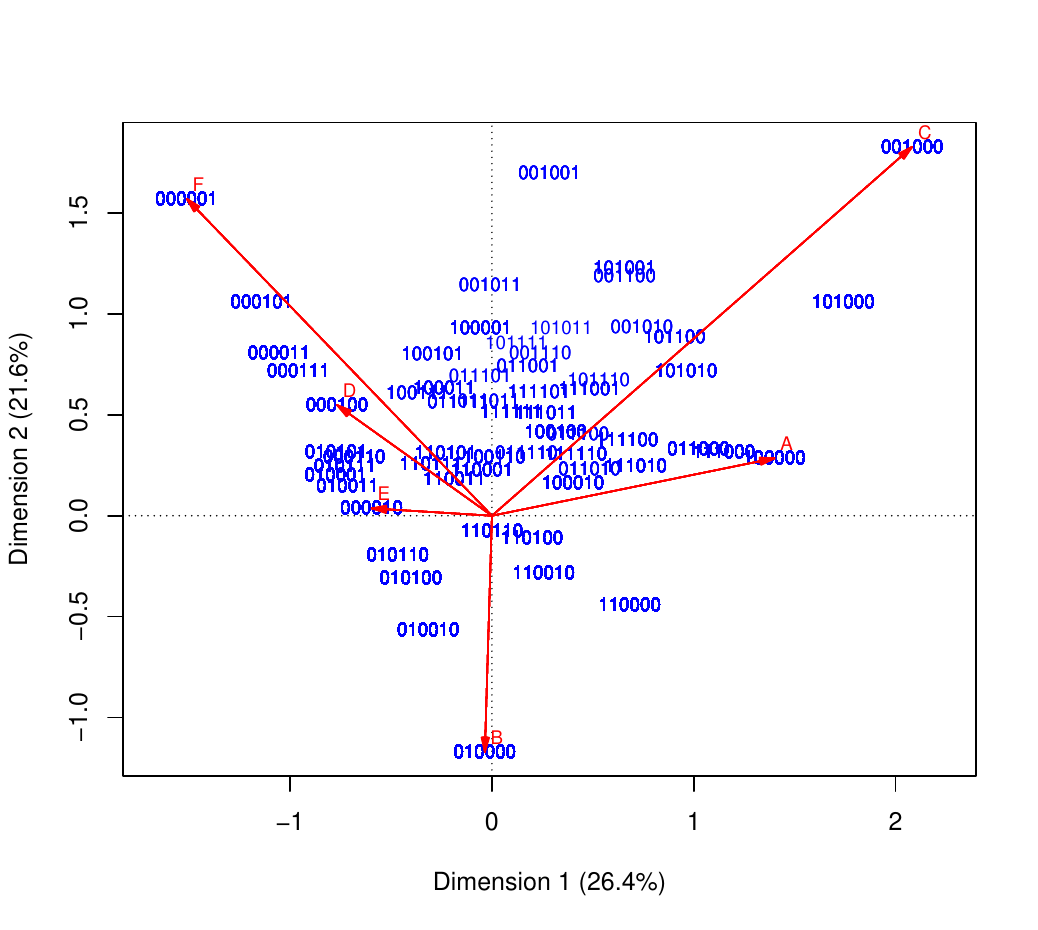}
    \caption{Two dimensional Correspondence Analysis solution for Religious practices data.}
    \label{fig:sugi_compare}
\end{figure}

Heiser (1981, pp. 142 - 143)  analysed these data using  Correspondence Analysis. In CA, a researcher has to choose between the row-principal, column-principal, or symmetric normalization. Heiser opted for the row-principal normalization, such that the answer patterns are in the centroid of the item points. We reproduced the analysis in the Figure \ref{fig:sugi_compare}. 
We focus on the unimodal model approximation (i.e., distance rule) as suggested by \cite{terbraak1985correspondence}, but sometimes also discuss the standard inner product interpretation (i.e., projection rule). The first two dimensions explain 26.4 and 21.6 percent of the inertia, respectively. The column masses are 0.18, 0.35, 0.06, 0.13, 0.17, and 0.11, for items A till F respectively. 

The solution is in some aspects similar to ours. The order of the items in the CA solution along the V-shape from upper left, to bottom center, to upper right is FDEBAC. This same ordering can also be witnessed from our solution (Figure \ref{fig:sugi}) along a nonlinear curve, although the spacing in our solution is different. Furthermore, in both solutions the items B, E, and D are close together, as well as A and C. 

Inspecting the correspondence analysis solution in Figure \ref{fig:sugi_compare} in more detail, we can find some interesting things. In the row principal normalization as depicted here, the points for positions of participants are in the center of the item positions chosen by the participant. Therefore, participants with response pattern 000001 are exactly on top of the item point for item F (the last one), and participants with response pattern 000101 are exactly in the middle of items D and F. The usual correspondence analysis interpretation uses the projection rule. This projection rule states that we project the participant point onto the vector and we multiply the length of the vector times the distance from the origin towards the projection point. Following this rule, we can conclude that these participants (i.e., those with response pattern 000101) have a higher association with item F than with item D, because the length of vector F is longer. In contrast, following the distance model interpretation the two items are equally distant, which would result in this interpretation of equal association. The expected values for items D and F for those participants, are 0.64 and 0.95. The higher expected value cannot be explained from the distance perspective, as the distances are equal and the mass for item D is higher than the mass for item F. 

Inspecting the fitted values of this two dimensional correspondence analysis solution, we find values in the range -0.70
till 1.52, with values below zero and above 1. Therefore, we can not interpret these values as probabilities and the approximation of correspondence analysis to a unimodal logistic model, as discussed in Section \ref{sec:comparisons}, fails in that sense. This result is, of course, similar to the comparison of fitting a usual linear regression model and a logistic regression model to a binary response variable. 

What is unclear from the correspondence analysis solution is how to make the classification when we use the distance rule. Whereas, in our distance model it is clear whether the probability of a participant for an item is smaller or larger than 0.5 (i.e., whether a participant point falls within or outside a circle), in the correspondence analysis solution no such regions are available. 

We further like to remark that choosing a different normalization in the correspondence analysis visualization, does not alter the fitted values nor the association values using the projection rule, but it does change the distances. So, the interpretation in terms of a distance model alters depending on the normalization chosen. Whereas in CA the user has to specify a normalization, for our mapping, we do not have to make a choice between different normalizations as the between sets distances are optimized. Furthermore, our analysis includes the offsets ($m_r$) as circles in the display, whereas such effects are usually not displayed in CA. As a consequence, from our display we can immediately see whether the probability of endorsement is larger than 0.5 or not for a given subject point. Such information cannot be retrieved from the CA solution.

\subsection{Dutch Election Data}

This data set consists of 352 Dutch inhabitants and their vote intentions for the parliamentary election in 2002 \citep{dpes0203}. 
The responses in this data set correspond to vote intentions for 8 different political parties: PvdA  (110), CDA (123), VVD (119),  D66 (89),  GL (114),  LN (27),  LPF (77),  and SP (57).
Ninety-three respondents indicated only one party, 169 respondents reported a vote intention for two parties, 79 for three parties, 8 for four parties, 2 respondents for five parties, and 1 respondent still had 6 parties under consideration. 

Furthermore, respondents were asked their opinion on five issues. These opinion data can be used as predictors. On a seven point scale, they had to indicate whether they think that \emph{Euthanasia} (E) should always be forbidden (1) or that a doctor should always be allowed to end a life upon a patient’s request (7). Similarly, whether \emph{Income Differences} (ID) should be increased (1) or decreased (7). The third issue concerns \emph{Asylum Seekers} (AS), and whether the participants have the opinion that  the Netherlands should allow more asylum seekers to enter (1) or should send back as many asylum seekers as possible (7). The next issue is about the acting of the government towards \emph{Crime} (C), that is whether the government is acting too tough on crime (1) or  should act tougher on crime (7). Finally, the participants had to indicate their location on an 11-point \emph{Left-Right} (LR) scale, where 0 indicates left and 10 right wing. We centered these predictor variables around 4 for the seven points scales and around 5 for the left-right variable. 

\subsubsection{Unsupervised analysis}

\begin{figure}
\begin{center}
\includegraphics[width = .9\textwidth]{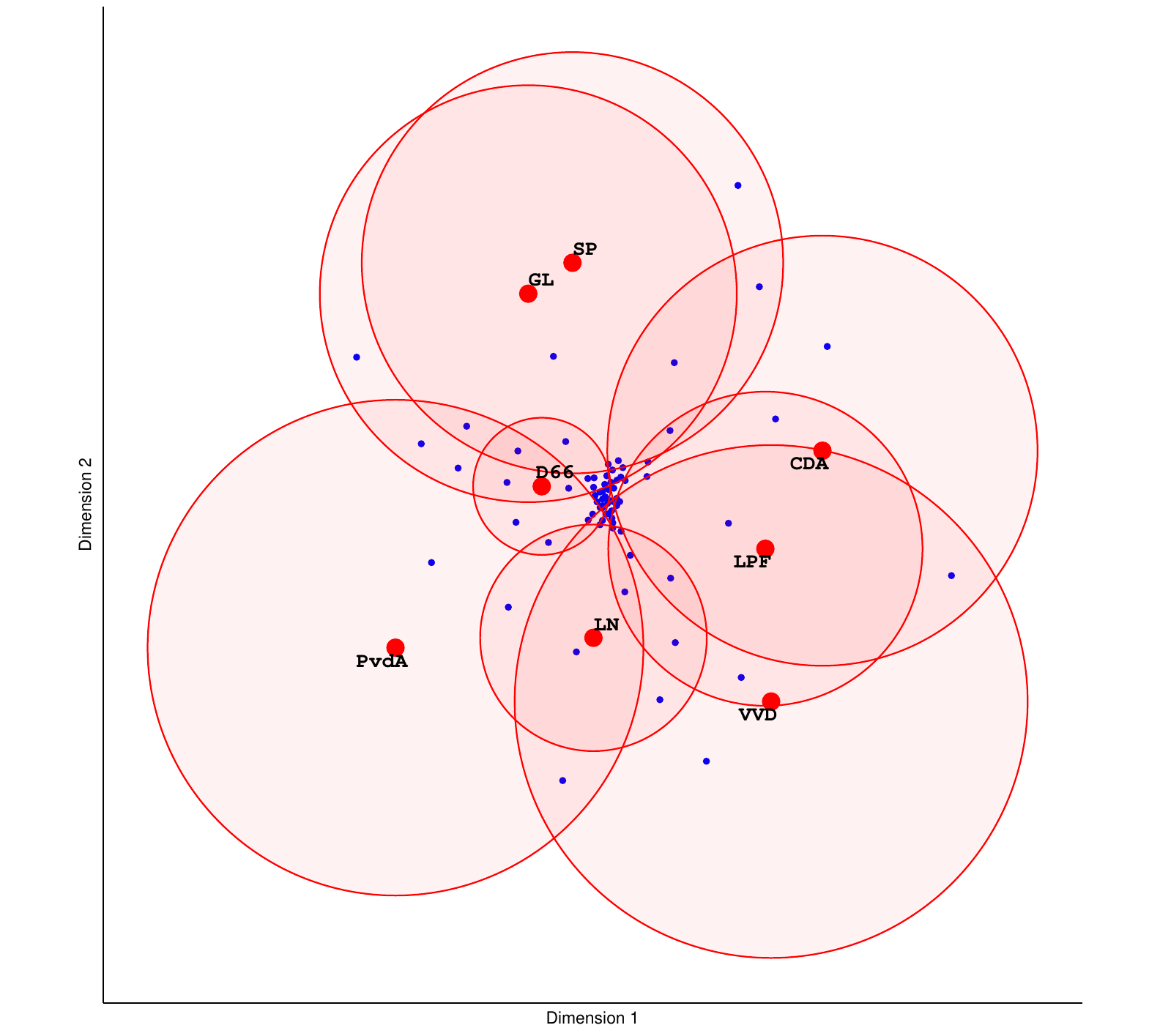}
\caption{Two dimensional unsupervised solution for Dutch Election data.}
\label{fig:dpes1}
\end{center}
\end{figure}

In this unsupervised analysis, we only consider the vote intention data as responses. The deviance of the intercepts only model for these data equals 3056.7. In this first analysis of the Dutch election data, we do not take the predictors into account, so we perform an unsupervised analysis. The optimal deviance for the unidimensional representation is 2,089.7 (31.6\% explained deviance), for the two-dimensional representation 918.52 (70.0\% explained), and for the three-dimensional representation 295.1 (90.0\% explained). 
The AICs are 2823.7, 2368.5, and 2459.1 for the one, two, and three dimensional solution, respectively. 
The two-dimensional map is displayed in Figure \ref{fig:dpes1}. 

Both PvdA and VVD have large areas of endorsement, while D66 has the smallest. We see that SP and GL are close together with large overlap in the region of endorsement, both are left-wing oriented parties. More right-wing oriented parties are VVD, LPF and LN; they group together on the other end. D66 and CDA are traditionally more centered parties, where CDA is a Christian party, while D66 is a progressive party. The PvdA is the Dutch labour party. 

From the solution, we computed some classification statistics for each political party: the proportion correctly classified, the sensitivity and specificity, the positive predictive value and negative predictive value, the F1-score and the Area under the ROC curve (AUC). See \cite{lever2016classification} for a definition and discussion about the use of these metrics. These statistics depend on the observed values $y_{ir}$ and the fitted values $\hat{\pi}_{ir}$. 
These statistics are shown in the top of Table \ref{tab:class12} and show that, generally, the representation is good. 
Note that, we evaluated the statistics \emph{in sample}, that is, the statistics are computed using the same sample that was used to fit the model. Although usually, these statistics would be evaluated \emph{out of sample}, the main reason for in sample evaluation is to compare these statistics for the unsupervised and supervised analysis in the next section. 


\begin{table}
\caption{Classification statistics for each political party based on the unsupervised solution and supervised solution.}
\centering
\begin{tabular}[t]{lrrrrrrrr}
\toprule
  & \multicolumn{8}{c}{\emph{Unsupervised analysis}}\\
  & PvdA & CDA & VVD & D66 & GL & LN & LPF & SP\\
\midrule
Proportion correct & 0.770 & 0.784 & 0.784 & 0.878 & 0.838 & 0.905 & 0.851 & 0.851\\
Sensitivity (TP / P) & 0.824 & 0.727 & 0.792 & 0.800 & 0.792 & 0.909 & 0.833 & 1.000\\
Specificity (TN / N) & 0.754 & 0.808 & 0.780 & 0.918 & 0.860 & 0.905 & 0.857 & 0.828\\
Pos Pred Value & 0.500 & 0.615 & 0.633 & 0.833 & 0.731 & 0.625 & 0.652 & 0.476\\
Neg Pred Value & 0.935 & 0.875 & 0.886 & 0.900 & 0.896 & 0.983 & 0.941 & 1.000\\
F1 & 0.622 & 0.667 & 0.704 & 0.816 & 0.760 & 0.741 & 0.732 & 0.645\\
AUC & 0.880 & 0.904 & 0.906 & 0.953 & 0.917 & 0.955 & 0.932 & 0.892\\
\midrule
  & \multicolumn{8}{c}{\emph{Supervised analysis}}\\
  & PvdA & CDA & VVD & D66 & GL & LN & LPF & SP\\
\midrule
Proportion correct & 0.679 & 0.662 & 0.699 & 0.750 & 0.764 & 0.923 & 0.807 & 0.841\\
Sensitivity (TP / P) & 0.475 & 0.533 & 0.587 & 1.000 & 0.707 & - & 0.696 & 0.538\\
Specificity (TN / N) & 0.722 & 0.688 & 0.729 & 0.749 & 0.780 & 0.923 & 0.815 & 0.853\\
Pos Pred Value & 0.264 & 0.260 & 0.370 & 0.011 & 0.465 & - & 0.208 & 0.123\\
Neg Pred Value & 0.868 & 0.878 & 0.867 & 1.000 & 0.908 & - & 0.975 & 0.980\\
F1 & 0.339 & 0.350 & 0.454 & 0.022 & 0.561 & - & 0.320 & 0.200\\
AUC & 0.679 & 0.697 & 0.743 & 0.618 & 0.786 & 0.768 & 0.749 & 0.748\\
\bottomrule
\end{tabular}
\label{tab:class12}
\end{table}

\subsubsection{Supervised analysis}\label{sec:dpes2}

In this subsection, we add the predictors to the analysis. The vote intention are the response variables, the opinions the predictor variables. For the supervised analysis, there is no need to remove the zero profiles from the analysis, as now the predictors have information for those participants. 

\begin{figure}
\includegraphics[width = 1.1\textwidth]{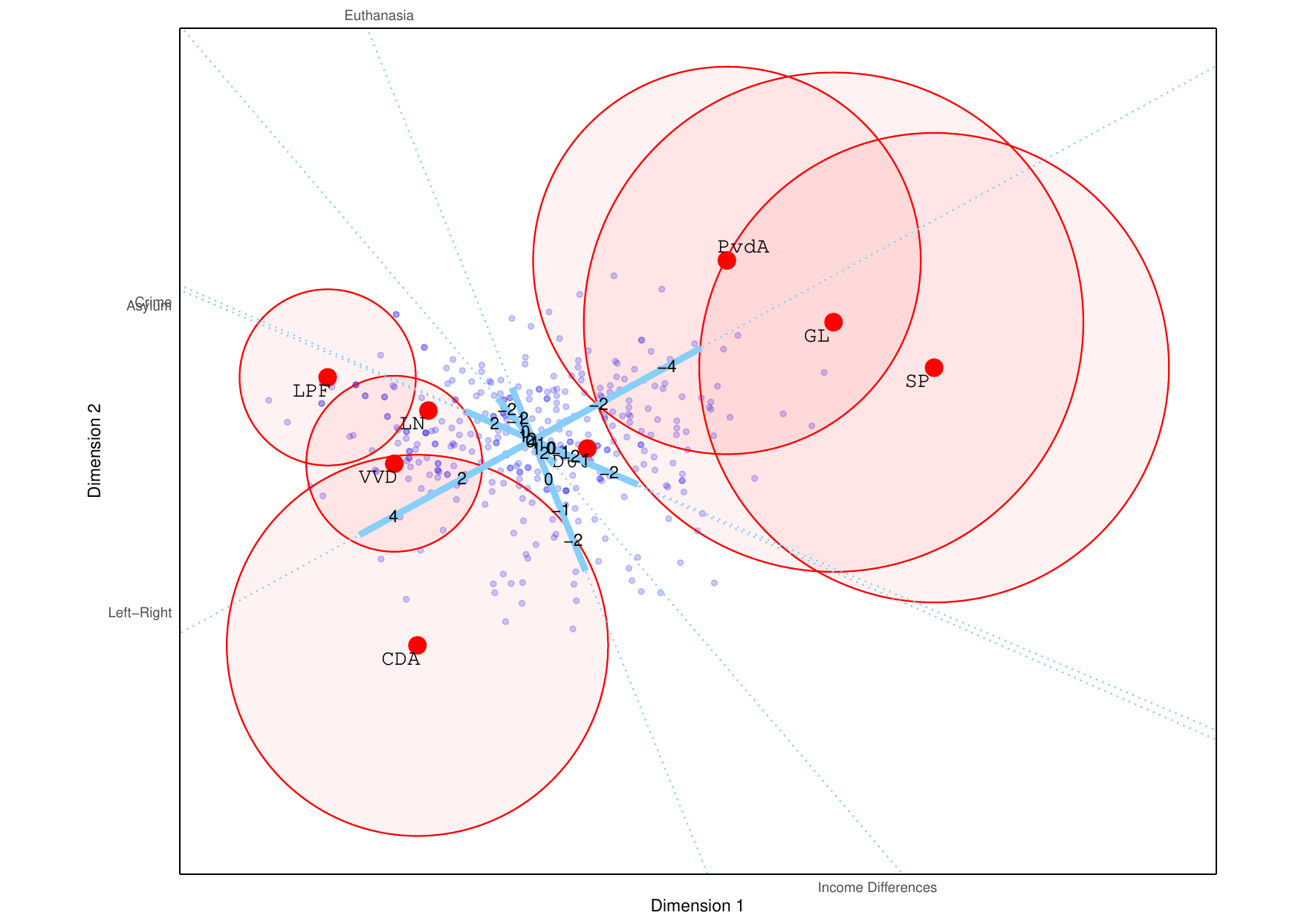}
\caption{Two dimensional supervised solution for Dutch Election data. The political parties (red points) represent the responses (i.e., intention to vote for these parties), the circles around these point represent regions of endorsement where the probability is larger than 0.5, the smaller dots represent the participants and the predictor variables are represented by variable axes with labels indicating the values of the predictor variable.}
\label{fig:dpes2}
\end{figure}

The optimal deviances are 2,770.2, 2,716.3, and 2,681.4 for the one, two, and three-dimensional solution respectively.  The two dimensional solution explains 11.1\% of the total deviance (i.e., compared to the intercept only model) and explains 15.9\% of the deviance of the unsupervised solution. 
The AICs are 2810.2, 2778.3, and 2763.4 for the one, two, and three dimensional solution, respectively. Although the AIC favors the three dimensional solution, we further discuss the two-dimensional solution in Figure \ref{fig:dpes2}, such that we can compare the supervised solution with the unsupervised one. 

Before, we delve into the interpretation, however, we further assess the quality of the mapping. Inspecting the deviance residuals (results not shown), we notice that for 5 participants these residuals are larger than twice the average. Having a closer look at these five observations shows no worrying signs, however. The contributions of the 8 response variables to the overall deviance are 15\% for PvdA, 15\% for CDA, 14\% for both VVD and D66), 13\% for GL, 6\% for LN, 12\% for LPF, and 10\% for SP. There is some variation in fit, but there are no political parties that have an extreme contribution to the overall deviance. We also assessed the influence of individual observations. Results are shown in Figure \ref{fig:influence} in Appendix A, where it can be seen that removal of observation 162 leads to a relatively large change in weights and class points, but not in the deviance. Removing this participant from the data did, however, not results in a large change of interpretation. Finally, we also checked the model specification. The component plus residual plots for our mapping are shown in Figure \ref{fig:compres} (Appendix A). Overall, the mapping seems to be well specified. For the predictor variable crime there seems to be some minor misspecification in the lower values, especially for response variables CDA, VVD, and LPF. We could try to improve the model specification, by including the quadratic term of crime. Note, however, that including such a quadratic term also influences the relationship of this predictor with the responses that seem to be well specified.   

Compared to the unsupervised analysis, in this supervised analysis variable axes are included for the predictor variables. The solid part of the lines correspond to values within the observed range (minimum to maximum) of the predictor variable, while the dotted lines extend the variable axis to the border of the display. The length of the solid part of the variable axes can be considered a type of effect size as it displays how much difference a specific predictor variable makes in the positions of the persons in the joint map. The higher the contribution of the variable to the scatter of the person positions, the larger the contribution. Variable names are printed at the positive side of the variable (i.e., high scores). The left-right variable for example, for which positive scores indicate that the subject considers him or herself right-wing, runs from upper right to lower left. Participants that consider themselves right wing are therefore located in the lower left quadrant of the space. Subject positions can be obtained from the variable axes by \emph{interpolation} or a process called completing parallelograms \citep[see][]{gower1996biplots, gower2011understanding}. 

Considering the positions, we see that PvdA is now close to the other two left-wing parties, SP and GL. The three right-wing parties (VVD, LPF, and LN) are also closer together. D66 is positioned more in the center, while the CDA stands out a bit. When participants consider themselves left-wing the probability for voting left-wing parties PvdA, GL, and SP is higher. Similarly, when participants consider themselves right-wing, the probability of voting one of the right wing parties (LPF, VVD, LN, or CDA) is higher. When participants indicate that the Netherlands should send back as many asylum seekers as possible and act tougher on crime, they have a higher probability of voting for LPF, LN, or VVD, while those participants that indicate the opposite have a high probability for voting either SP or GL.  Participants that indicate euthanasia should always be forbidden, have a higher probability for voting either CDA or SP. 

For this supervised analysis we computed the same classification statistics as for the unsupervised analysis (again, in sample). The results are shown in lower half of Table \ref{tab:class12}. Overall, we see that the classification evaluation metrics become less good compared to the unsupervised analysis. That is expected as we restrict the coordinates of the participants to be linear combinations of the issue opinions, that are the predictor variables. Because the estimated offset parameter for LN ($\hat{m}_{LN}$) is negative, the probability of choosing LN never exceeds 0.5 and therefore some statistics cannot be computed for this party. The positive predictive value and the F1-score are overall quite low compared to the unsupervised solution.


\section{Monte Carlo experiments}\label{sec:sims}

In this section, we report on two Monte Carlo experiments. The first considers parameter recovery, the second considers predictive performance. 

\subsection{Parameter recovery}
In this section, we will discuss numerical experiments investigating the ability of the algorithm to recover a population distribution. The population distributions are based on the empirical examples of the previous section. 

\subsubsection{Data generation}

In our experiments, we take the estimated parameters from Section \ref{sec:applications} and some characteristics of the data sets as population parameters. 

For the \emph{unsupervised analysis}, we take the mean and covariance matrix from the estimated subject positions and draw a population $\mathbf{U}$ of 100,000 subjects. The estimated response locations (from Figure \ref{fig:sugi} or \ref{fig:dpes1} ) are taken as the population parameters $\mathbf{V}$.
 
For the \emph{supervised analysis}, we draw 100,000 values from a multivariate normal distribution with mean zero and covariance matrix equal to the observed covariance matrix of the predictors in the Dutch Election data. With these generated predictor variables and the population regression weights ($\mathbf{B}$), we compute subject locations ($\mathbf{U}$). 

For both unsupervised and supervised analysis, we calculate probabilities using the distances between the person locations and the item locations and using the estimated offsets ($m_r$) as population values. In every replication, we draw a subsample from the persons and use the probabilities ($\pi_{ir}$) to draw responses variables ($y_{ir}$) from the binomial distribution. The subsamples have different sample sizes, that is $n = 100$, 200, 500, and 1000. We ﬁt the mapping on the generated data. One-hundred replications are used.

\subsubsection{Evaluation}

Let a population configuration be defined by the matrix $\mathbf{Z}$ and its estimate as $\hat{\mathbf{Z}}$. 
The \emph{congruence coefficient} is a measure of recovery of the population configuration and is defined as \citep[][p. 350]{borg2005modern}
\begin{equation}
\label{eq:congr}
\phi = \frac{\sum_{i<j} d_{ij}(\mathbf Z)d_{ij}(\hat{\mathbf Z})}{\sqrt{\sum_{i<j} d_{ij}^2(\mathbf Z)}\sqrt{\sum_{i<j} d_{ij}^2(\hat{\mathbf Z})}}.
\end{equation}

We define two such measures, the first $\phi_{uv}$ for the complete configuration including subject and item points, such that $\mathbf{Z} = \left[ \mathbf{U}^\prime, \mathbf{V}^\prime \right]^\prime$, whereas the second $\phi_v$ only uses the item points, that is $\mathbf{Z} = \mathbf{V}$.

Another measure of configuration similarity is obtained by computing the \emph{product-moment correlation coefficient} over the coordinate matrices $\mathbf{Z}$ and $\hat{\mathbf{Z}}$, defined as \citep{borg2022note}
\begin{equation}
\label{eq:corr}
r = \frac{\mathrm{tr}(\mathbf{Z}^\prime \hat{\mathbf{Z}}\mathbf{T})}{\sqrt{\mathrm{tr}(\mathbf{ZZ}^\prime) \cdot \mathrm{tr}(\hat{\mathbf{Z}}\hat{\mathbf{Z}}^\prime)}},
\end{equation}
where $\mathbf{T}$ is the estimated Procrustean rotation matrix, and $\mathrm{tr}(\mathbf{A}) =  \sum_{i} a_{ii}$. For evaluation of recovery of the population configuration we focus again on the complete configuration with subject and item points ($r_{uv}$) and only the item point configuration ($r_v$)

\subsubsection{Results for Unsupervised Mapping}\label{sec:simun}

The simulation results for the Monte Carlo study based on the Sugiyama data and on the election data are given in Table \ref{tab:simdpes1}. The recovery of the population configuration is overall very good for both studies. The recovery becomes better with larger sample sizes, that is the mean values increase while the standard deviations decrease. Both observations are especially visible for $\phi_v$ and $r_v$.

\begin{table}[ht]
\caption{Results from the simulation study for unsupervised analyses.} 
\centering
\begin{tabular}{llr|rrrr}  \hline
		&			&		& \multicolumn{4}{c}{Sample Size} \\
Data Set  	& Measure	& 		& 100    & 200    & 500    & 1000 \\  \hline
Religious	& $\phi_{uv}$	& mean 	& 0.948 & 0.948 & 0.952 & 0.955 \\ 
  		& 			& std 	& 0.007 & 0.008 & 0.006 & 0.003 \\ 
  		& $\phi_{v}$ 	& mean	& 0.976 & 0.979 & 0.987 & 0.993 \\ 
  		& 			& std		& 0.015 & 0.011 & 0.008 & 0.003 \\ 
		&  $r_{uv}$	& mean	& 0.889 & 0.889 & 0.896 & 0.900 \\ 
   		& 			& std		& 0.018 & 0.015 & 0.011 & 0.005 \\ 
		&  $r_{v}$ 		& mean	& 0.952 & 0.957 & 0.974 & 0.986 \\ 
   		& 			& std 	& 0.028 & 0.021 & 0.015 & 0.006 \\ 
   \hline
Election	& $\phi_{uv}$   	& mean 	& 0.949 & 0.949 & 0.948 & 0.947 \\ 
  	 	&			& std 	& 0.007 & 0.005 & 0.003 & 0.002 \\ 
		& $\phi_{v}$	& mean 	& 0.982 & 0.991 & 0.996 & 0.997 \\ 
 	 	&			& std 	& 0.009 & 0.005 & 0.002 & 0.001 \\ 
		&  $r_{uv}$ 	& mean	& 0.893 & 0.896 & 0.899 & 0.898 \\ 
  	 	& 			& std 	& 0.014 & 0.009 & 0.005 & 0.004 \\ 
		&  $r_{v}$ 		& mean	& 0.961 & 0.979 & 0.990 & 0.992 \\ 
  	 	& 			& std 	& 0.018 & 0.010 & 0.003 & 0.002 \\ 
   \hline
\end{tabular}
\label{tab:simdpes1}
\end{table}

\begin{figure}[ht]
\noindent\begin{subfigure}[b]{0.5\textwidth}
\includegraphics[width = .95\textwidth]{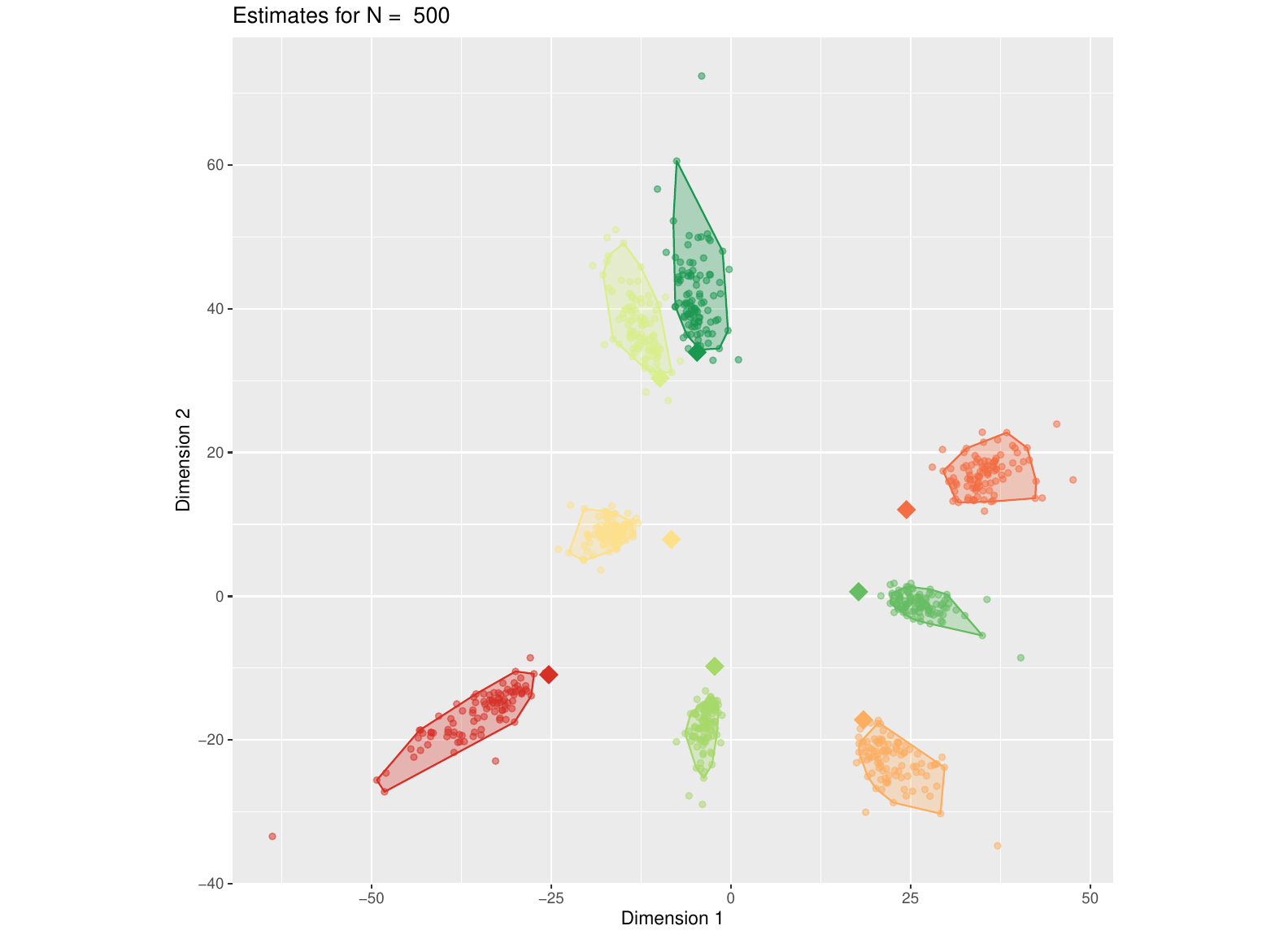}
\caption{}
\label{fig:procrustes1}
\end{subfigure}%
\noindent\begin{subfigure}[b]{0.5\textwidth}
\includegraphics[width = .95\textwidth]{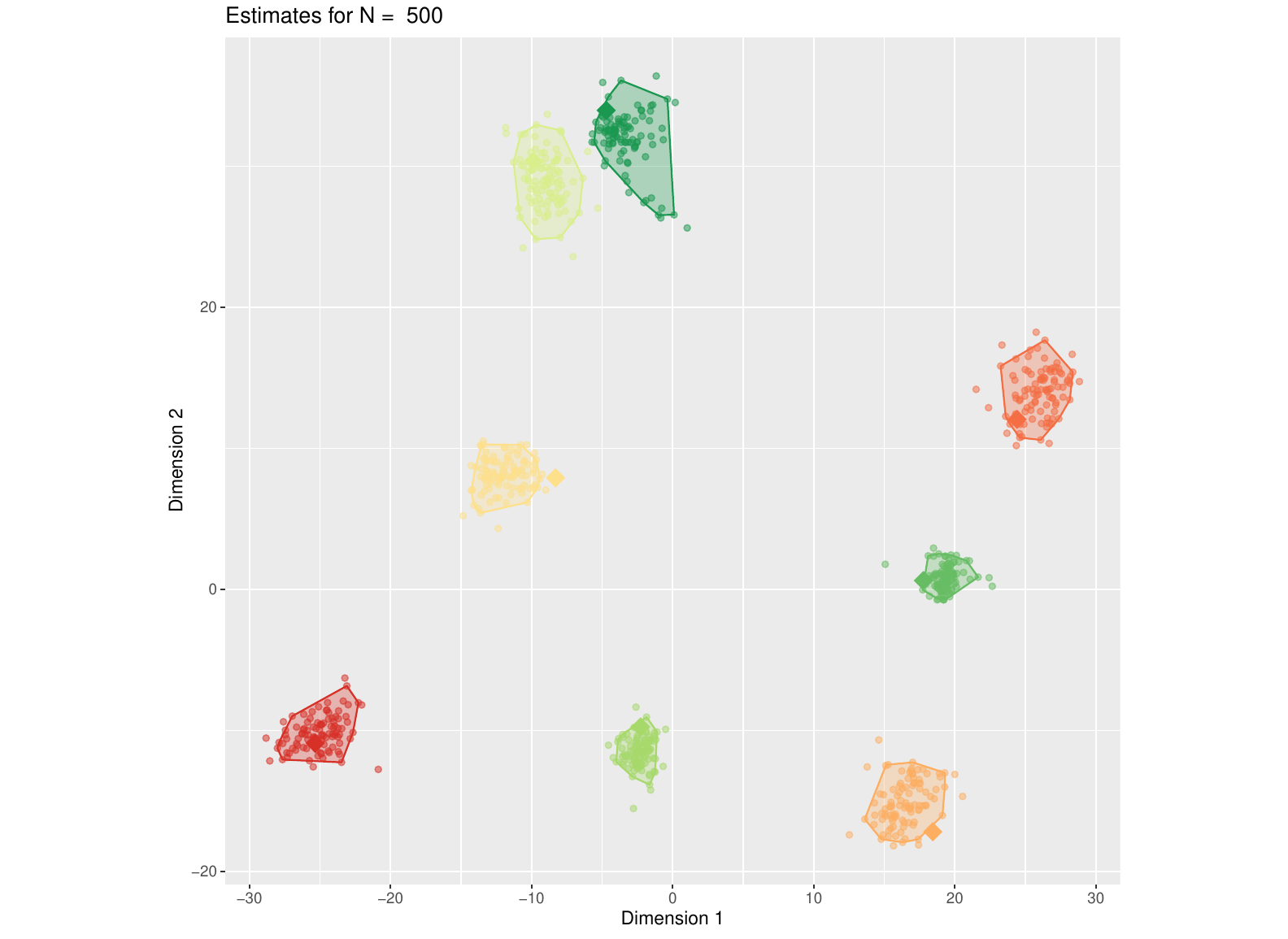}
\caption{}
\label{fig:procrustes2}
\end{subfigure}%
\caption{Graphical representation of the results of the simulation study based on the politics data with sample size equal to 500. Panel (a) presents the results after an orthogonal Procrustes analysis without dilation; Panel (b) after a similarity transformation with dilation. The diamonds show the population points, the dots show the estimated values. The bags include 90\% of the estimated locations.}
\label{fig:procrustes}
\end{figure}

Although the recovery is very good according to these statistics, we need to be somewhat cautious. The congruence coefficient and the correlation coefficient assume that the geometric structure may be translated, dilated, and rotated. Our mapping method, however, only has translational and rotational freedom. Therefore, it is informative to geometrically show the estimated configurations after 1) an orthogonal Procrustes analysis towards the population distribution, and 2) a Procrustes similarity analysis, which also takes the dilation into account. In Figure \ref{fig:procrustes}, we show the results of the estimated configurations after orthogonal Procrustes (left) and after a similarity transformation (right). The diamonds show the population points, the dots show the estimated values. The bags include 90\% of the estimated locations. Whereas in the similarity Procrustes the population values are generally within the 90\% bags, for the orthogonal Procrustes analysis we see that estimated positions are more extreme (further away from the origin) than the population values and the population points often fall outside the bags. This feature represents a kind of overfitting, that is, with larger distances and simultaneously larger estimated $m_r$ parameters, the algorithm produces estimated probabilities closer to zero and one. For minimizing the negative log-likelihood this is advantageous. 


\subsubsection{Results for Supervised Mapping}

The recovery results for the supervised algorithm are shown in Table \ref{tab:simdpes2}, where it can be seen that the recovery is very good for all sample sizes. The mean values of the coefficients increase with sample size and the standard deviation of the congruence and correlation coefficients decrease with larger sample size. In contrast to the results of the unsupervised case, in this case we found no signals of overfitting. 

\begin{table}[ht]
\caption{Simulation results for Supervised Algorithm}
\centering
\begin{tabular}{lr|rrrr}  \hline
			&		& \multicolumn{4}{c}{Sample Size} \\
Measure		& 		& 100    & 200    & 500    & 1000 \\  \hline
$\phi_{uv}$  	& mean 	& 0.956 & 0.980 & 0.994 & 0.997 \\ 
		  	& std 	& 0.020 & 0.009 & 0.003 & 0.002 \\ 
$\phi_{v}$  	& mean 	& 0.954 & 0.980 & 0.994 & 0.998 \\ 
		  	& std 	& 0.029 & 0.015 & 0.004 & 0.001 \\
$r_{uv}$  		& mean 	& 0.908 & 0.958 & 0.986 & 0.994 \\ 
			& std 	& 0.047 & 0.019 & 0.006 & 0.003 \\ 
$r_{v}$ 		& mean 	& 0.896 & 0.952 & 0.986 & 0.995 \\ 
	 		& std  	& 0.062 & 0.030 & 0.009 & 0.002 \\ 
\hline
\end{tabular}
\label{tab:simdpes2}
\end{table}

\subsection{Predictive Performance}

In this second Monte Carlo study, we compare the predictive performance of our logistic multidimensional unfolding with logistic reduced rank regression. Both methods are comparable in the sense that they produce a low dimensional mapping of the data with approximately the same number of parameters. In Section \ref{sec:probs} we showed that multidimensional unfolding can represent a larger number of response profiles in the same dimensional space. The number of represented profiles increases as the offset parameters become larger because then the circles, representing the regions of endorsement, overlap. We expect that this leads to better predictive performance. 

\subsubsection{Data generation}

We generate data with $P = 3$ predictor variables from a multivariate standard normal distribution with uncorrelated predictors. The population parameters $b_{ps}$ are (1,0) for the first predictor, (0,1) for the second, and $(\sqrt{2}, \sqrt{2}$ for the third. We use $R = 13$ response variables, with the following coordinates in two-dimensional space
\[
\mathbf{V}^\prime = 
\left[ \begin{array}{rrrrrrrrrrrrr}
1 & \frac{1}{2} & \frac{1}{2} & \frac{1}{2} & 0 & 0 & 0 & 0 & 0 & -\frac{1}{2} & -\frac{1}{2} & -\frac{1}{2} & -1 \\
0 & \frac{1}{2} & 0 & -\frac{1}{2} & 1 & \frac{1}{2} & 0 & -\frac{1}{2} & -1 & \frac{1}{2} & 0 & -\frac{1}{2} & 0
\end{array}\right]
\]
Values for the offsets $m_r$ are drawn in every replication from a uniform distribution. We consider three ranges, the first from -1 till 0, the second range is from -0.5 till 0.5, and the third range from 0 to 1. With these settings, we generated training data sets with 200, 500, and 1000 observations from either the distance model or the inner product model.
We also generated for each training set a test set with 1000 observations. We used 100 replications. 

\subsubsection{Evaluation}

Each generated training data set is analysed by both a logistic restricted multidimensional unfolding and a logistic reduced rank regression. When the data are generated by the distance model, we start logistic multidimensional unfolding with the population parameters as initial values and trust this will give good results. When the data are generated by the inner product model, we start logistic multidimensional unfolding with the population parameters as initial values but also perform 25 random starts. The solution with the lowest negative log-likelihood is saved. Logistic reduced rank regression is not hampered by the local optima problem, so for this procedure starting values do not matter. 

With the estimated parameters of these two procedures and the predictor variables in the test set, we compute predicted values \(\hat{\pi}_{ir}\) for the observations in the test set. We compare the predictive performance of the two methods using the Brier score
\[
\sum_{i = 1}^{1000} \sum_{r = 1}^{13} (y_{ir} - \hat{\pi}_{ir})^2 / 13000,
\]
and compare them using boxplots. 

\subsubsection{Results}

\begin{figure}
\begin{center}
\includegraphics[width = .9\textwidth]{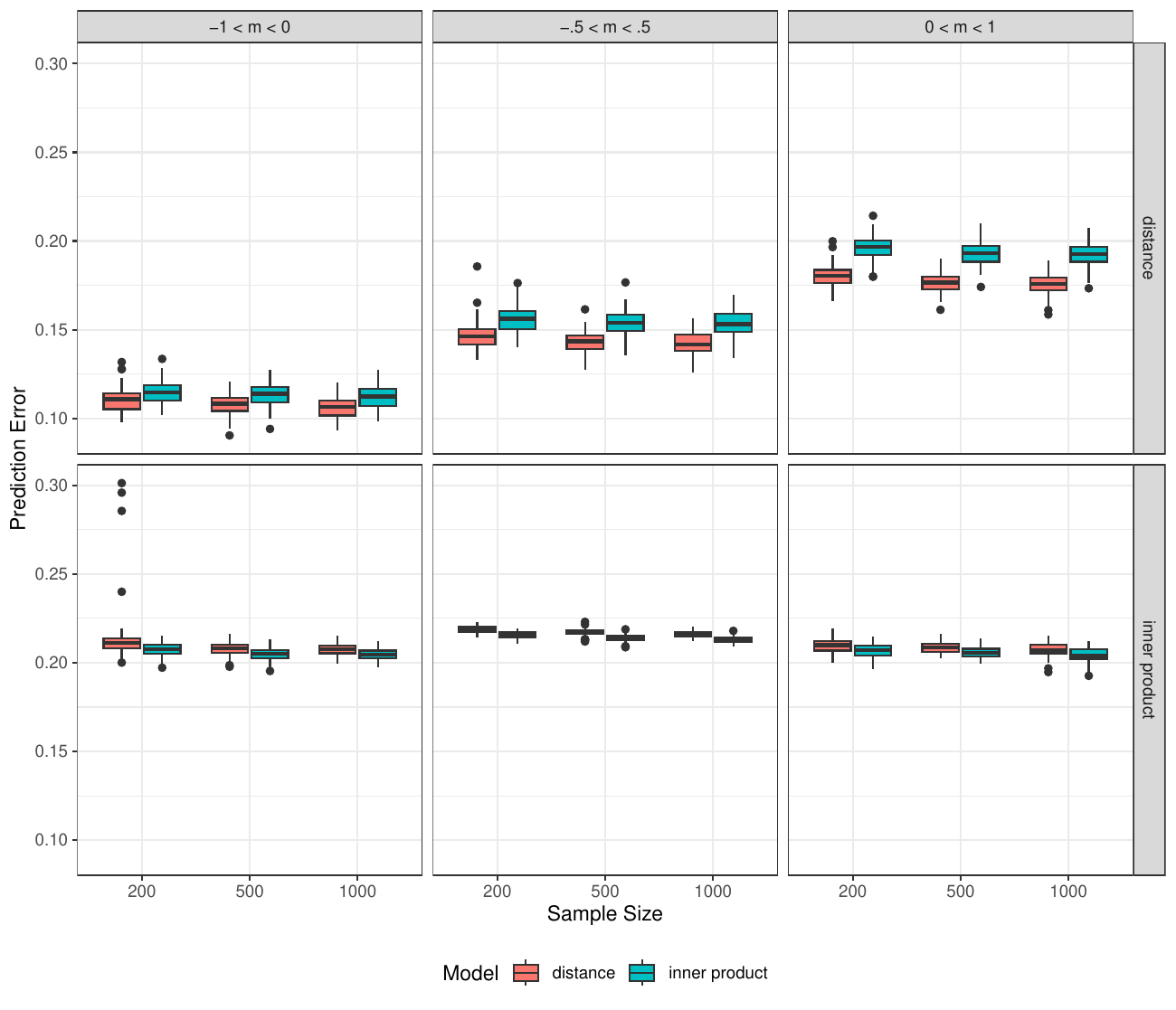}
\caption{Prediction error for restricted logistic multidimensional unfolding (distance, left boxplot) compared to logistic reduced rank regression (inner product, right boxplot) for data generated with both population models. In the upper row, data are generated following the distance model whereas in the lower row data are generated with the inner product model. The three different columns represent different ranges of population offset parameters.}
\label{fig:predsim}
\end{center}
\end{figure}

The results of this simulation study are shown in Figure \ref{fig:predsim}, where the upper row shows the results for data sets generated with the distance model, whereas the lower row shows the results for data sets generated with the inner product model. The upper and lower row cannot simply be compared because the characteristics of the data, such as the proportion of ones for each response variable differ. The focus therefore is on the predictive performance of the distance model as compared to the inner product model. The three columns in Figure \ref{fig:predsim} correspond to the different ranges from which the offsets are drawn. 

In case the data are generated with the distance model (upper row), we see that the distance model predicts the test data better than the inner product model. The differences between the two become larger with higher values of the offsets. The reason is that with higher offset values the regions of endorsement become overlapping, leading to a larger number of possible response patterns. The distance model can accommodate a larger number of response patterns than the inner product model.

When the data are generated from a inner product model (lower row), the inner product model predicts the test data slightly better than the distance model. No matter what the values for the offset parameters are (i.e., comparing the left, middle, and right columns of plots), the \emph{difference} in predictive performance of the two models is approximately the same. As the response curves of the inner product model can be considered a special case of those of the distance model (i.e., take only part of the response curve till (or from) the peak of the curve), we did not expect the differences to be large. The slightly better predictive performance of the inner product model is probably due to an overfitting effect of the distance model, that is, the positions of the response variables are on the boundary of the solution with large offsets, so that the decision boundaries resemble the straight lines of the reduced rank model. Furthermore, the distance model uses a few more parameters, generally leading to an increase in variance and therefore possibly worse predictions. The difference between the two sets of predictions are, however, really small. 

In the left lower plot, we see a few outliers with high prediction errors for the distance model with small sample size. This is probably due to a local optimum and could be resolved by more random starts. 

\section{Discussion}

We proposed a new mapping methodology for binary variables based on the two-mode distance function. The distances in the map are connected through a logistic function with the probabilities or expected values. The distinctive feature of the mapping is that it assumes a proximity answer process, where the distance between a person point and an item point determines the probability to endorse or agree with an item: A small distance corresponds to a large probability, whereas a large distance corresponds to a small probability. Every item or response variable is represented by a point and a circle. Distances towards the point are directly related to probability of endorsement. The circles define a region of endorsement, that is, a region where the probability of endorsement is larger than 0.5. When an estimated offset parameter is negative, the corresponding item does not have a region of endorsement because the probability of endorsing never exceeds 0.5. In that case, for each participant we would predict that this item is not endorsed. However, for 100 participants each having a probability of endorsement of 0.3 we expect 30 of these participants to endorse the item. 

The proposed mapping can be used for supervised and unsupervised analyses. In a supervised analysis a set of explanatory or predictor variables is available for the participants that restrict the estimates of the points of the participants to be functions of these predictor variables. By including predictors in the map, we specify a functional form of the relationship between the predictors and the response variables. To check for misspecification, we generalized component plus residual plots that are sometimes used in generalized linear models. The plots are applied in the supervised analysis shown in Section \ref{sec:dpes2}. In the current paper, we only used additive linear functions but nonlinear and non-additive functions can simply be incorporated by, for example, using spline bases and/or interaction terms. Such terms, would alter the relationship between the variable axes and the participant points and depending on the precise specification might become difficult to interpret. The relationship between the positions of the participants and the item locations, however, remains the same. 

We theoretically compared our mapping to proximity models developed earlier by \cite{desarbo1986simple, desarbo1987constructing} and \cite{takane1998choice}. Unfortunately, no software is available anymore for these methods. The methods are similar, but these methods use squared Euclidean distances, where we use the distances itself. Usually, researchers interpret distances when looking at a map, not squared distances. \cite{desarbo1986simple, desarbo1987constructing} use offsets for persons, whereas we use offsets for items. Using offsets for persons greatly increases the number of parameters to estimate. In the algorithm section, we showed how offsets for persons could be estimated. We did not incorporate that in our software (yet). In theory it is possible to use both person and item offsets, however this would make the interpretation more difficult and further interpretational guidelines for this case need to be developed. For example, what would it mean if the circles of a person and an item overlap?

An MM-algorithm is proposed for estimation of the map. The main majorization step transforms the negative log-likelihood to a least squares function as shown by \cite{groenen2003} and \cite{deleeuw2006principal}. We extended this majorization step with weights for the response profiles. In the inner loop a weighted least squares multidimensional unfolding has to be performed, which can be done by the SMACOF algorithm. The dissimilarities in this step are, however, not guaranteed to be positive. Therefore, we adopted an idea of \cite{heiser1991generalized} to the unfolding situation. This step is again an MM algorithm, so that the algorithm is a \emph{double MM} algorithm.  MM algorithms generally have a linear rate of convergence \citep{hunter2004tutorial}, that is, they often need many iterations to converge. On the other hand, the computations within the iterations are usually quite simple. As we have a double MM algorithm, where we majorize the majorization function of the negative log-likelihood, the algorithm is slow. \cite{heiser1995convergent} discussed ways to increase the speed of MM algorithms. 
 
We applied the mapping to two empirical data sets. The first data set is about religious practices. We applied our mapping and compared the results against results obtained by \cite{heiser1981unfolding} using correspondence analysis, a least squares mapping technique for single peaked data. Correspondence analysis can be interpreted using a proximity perspective, as shown by \cite{terbraak1985correspondence}. We showed that such an interpretation is problematic. Our solution has a clearer link between the geometric structure and the probabilities of endorsement. In correspondence analysis, participants are in the center of the items they endorse, but there is not a direct function that translates distances to expected values or probabilities. 


The second data set is about vote intentions, where not only the intentions (yes or no) are available but we also have opinions of the participants on a set of opinions, that can serve as explanatory variables. We showed both the unsupervised and supervised analysis and compared the results in terms of mapping but also in terms of classification diagnostics. Including explanatory variables constraints a set of points and therefore classification statistics become worse. A supervised analysis, however, shows relationships between opinions and vote intentions, or more generally between explanatory variables and responses like in regression models. In that sense, our mapping is similar to multiple logistic regression models where we assume a single peaked relationship between the predictors and the responses. The dependencies between the different response variables are ``modelled'' by using a low-dimensional map. If we can verify the assumption that given the low dimensional relationship the responses are independent we could extend the mapping with likelhood-based statistics for model selection. 

We performed two Monte Carlo experiments. In the first experiment, we evaluated the algorithm and saw that the unsupervised analysis leads to \emph{overfitting}. Consider, for example, the factitious configuration in Figure \ref{fig:2d} and assume that the two response profiles that are not represented in the visualization are also not observed. In that case, the profiles would all fall in the correct region, that is, for each $y_{ir} = 1$ the corresponding probability ($\hat{\pi}_{ir}$) is larger than 0.5, and for each $y_{ir} = 0$ the corresponding probability is smaller than 0.5. Making the complete visualization twice as large, that is multiplying $\mathbf{U}$, $\mathbf{V}$, and each $m_r$ by two, changes the probabilities such that probabilities larger than 0.5 become even larger and probabilities smaller than 0.5 become even smaller. This makes the negative log-likelihood smaller. Therefore, the algorithm will make the map larger and larger, without really changing the overall appearance of the configuration. In empirical data analysis, not all response profiles are correctly represented and these incorrect profiles act as a counterforce against blowing up the map. Similar overfitting issues are also found for the logistic PCA \citep{song2019principal}. Possible solutions for this overfitting can be found by including a penalty in the optimization function, for example on the sum of squares of $\mathbf{U}$ or $\mathbf{V}$. In the supervised analysis, we did not find any signs of overfitting, so including predictor variables is also a solution. 

In the first Monte Carlo study we used the congruence coefficient and the product-moment correlation as outcome measures. Both measures are generally large even for random data \citep{borg2022note} and unfortunately no clear guidelines are available for interpretation. Another measure is the \emph{alienation} coefficient which is simply $\kappa = \sqrt{1 - \phi^2}$. The alienation yields values that vary over a greater range and might be easier to distinguish. All methods have been used mainly for multidimensional scaling with a relatively small set of objects. How these coefficients behave for multidimensional unfolding and our mapping needs further investigation. 

In the second Monte Carlo experiment, we evaluated the predictive performance of our mapping against that of a logistic reduced rank model. Data were generated from two population models, one based on inner products the other based on distances. We showed that when the data are generated with a distance model, the distance model has better predictive performance than the reduced rank model. The difference in predictive performance increases as the values of the offsets increase. With larger offsets the number of regions of endorsement increases, each corresponding to a response profile. As the distance model can represent a larger number of response profiles than the inner product model this was an expected result. When the data are generated using an inner product population model, the reduced rank model predicts slightly better than the distance model. The difference does increase of decrease with changing offsets. The difference in performance might be explained by overfitting. The distance model is capable of fitting monotone response patterns by moving the position of the item on the boundary of the configuration and creating a large offset. This can be inferred from Figure \ref{fig:probs}, where the curves are monotonically increasing from the left till the position of the item or decreasing from the point of the time to the right. The distance model uses slightly more parameters than the reduced rank approach because of different identification issues. In our unsupervised analysis we have translational and rotational freedom, which amounts to $S(S+1)/2$ and $S(S-1)/2$ indeterminacies, respectively. In the corresponding inner product model, the number of indeterminacies is $S^2$. More parameters usually leads to higher variance and reduced predictive accuracy.   

In our mapping we use the two-mode distance function. Not all two-mode distance functions give rise to a proximity answer process. In spatial voting models for roll call data \citep{poole1985spatial, clinton2004statistical, poole2011scaling} and in the MELODIC family \citep{derooij2023melodic2} also a two-mode distance function is used. The crucial difference between our mapping and these approaches is that in the latter the distance is defined between a subject and a category of an item, whereas in the current paper it is the distance between a subject and an item. When the two-mode distance function is defined towards categories of a binary item or response variable, the structure implies a dominance answer process, such as in (logistic) principal component analysis.   

Our unsupervised mapping in the unidimensional case ($S =1$) is similar to so called single-peaked item response models \citep{andrich1988application, hoijtink1990latent, andrich1993hyperbolic, roberts2000general}. 
In these item response models often the person parameters (our $\mathbf{u}_i$) are random effects and the focus is on creating a good measurement scale. Our mapping is not focussed on measurement per se, although in the unidimensional case it could be used for that.  Our supervised mapping in the unidimensional case is similar to an explanatory multidimensional single-peaked item response model. Explanatory item response models  include predictor variables for either the participants or items. While for the dominance answer process there have been quite some developments \citep{deboeck2004explanatory}, for single-peaked response processes explanatory variants are largely lacking although there has been some recent work on explanatory variants of the generalized graded unfolding model \citep{joo2022explanatory, usami2011generalized}. All these single-peaked item response models assume a unidimensional latent trait, whereas our mapping is multidimensional. For (single-peaked) item response models often many items are needed while our mapping can deal with a small number of items.

To conclude this paper, the supervised mappings we developed in this manuscript are solutions to what computer scientists call the \emph{multi-label classification} problem \citep{gibaja2014multi, herrera2016multilabel}. Each of the observations is characterised by a set of labels, that is, the set of responses for which $y_{ir} = 1$. \cite{gibaja2014multi} discuss three different tasks for multi-label learning: the label ranking task, the multi-label classification task, and the multi-label ranking task. The latter task generalizes the first two and provides at the same time a ranking of the labels as well as a bipartition for each observation. The approach we develop in this paper, as well as logistic reduced rank regression, is also a multi-label ranking task. \cite{gibaja2014multi} and \cite{herrera2016multilabel} discuss three ways of learning from multi-label data set: the data transformation approach, the method adaptation approach, and the ensemble of classifiers. Furthermore, \cite{gibaja2014multi} describe an overview of transformation and adaptation methods. The approach that we develop in this paper is a method adaptation approach. \cite{siblini2019review} give a review on dimension reduction in multi-label classification. They discussed that both the feature space (i.e., the predictors) as well as the labels space (i.e. the outcomes) can be reduced. Often the dimension reduction is performed independent of the classification. We present a dimension reduction approach that takes into account the multi-label classification task. That is, our loss function is targeted towards classification performance but the mapping finds a \emph{joint space} in which we embed points for the response variables, such that classification performance is optimized. This joint space lies within the column space of the predictor space. An issue in multi-label classification is label dependence, the correlation among labels. In the approach we developed, the dependency between labels is explicitly taken into account through the dimension reduction.


\section*{Appendix A: Model Assessment of Supervised Map for Dutch Election data}

\begin{figure}[t]
\begin{center}
\includegraphics[width = 0.8\textwidth]{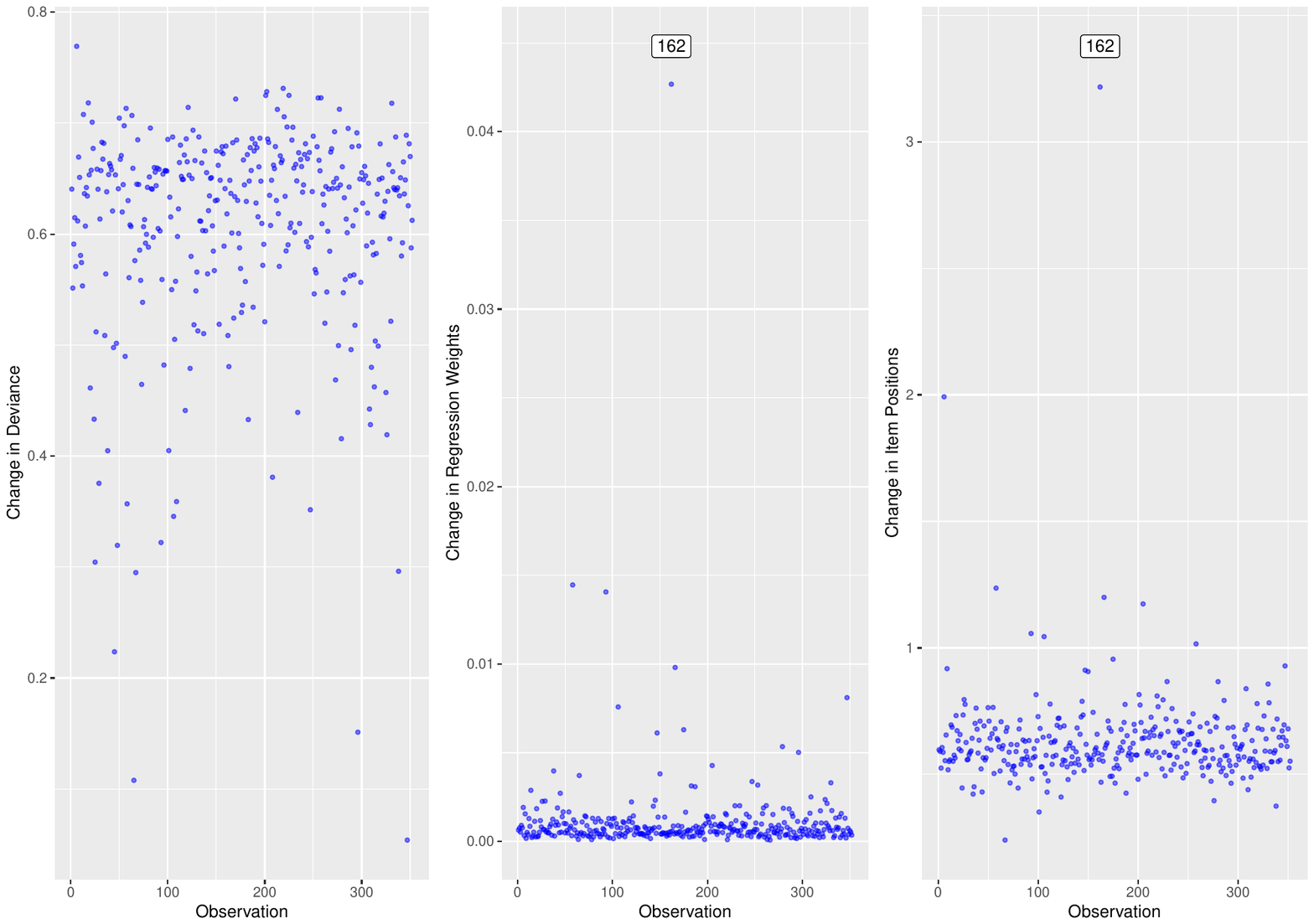}
\caption{Influence plots. Observation number (horizontal axis) against Change in deviance $\delta_{\mathcal{D}}$ (left), change in regression weights $\delta_B$ (middle), and change in item positions $\delta_V$ (right) on the vertical axis. Observation 162 has large values for change in regression weights (middle plot, point at the top) and change in item position (right plot, point at the top).}
\label{fig:influence}
\end{center}
\end{figure}

In Figure \ref{fig:influence}, we show the change in deviance, change in regression weights, and change in item locations statistics when deleting an observation for our analysis in Section \ref{sec:dpes2}. It seems that participant 162 has large influence as the regression weights and item locations change substantially when deleting this observations. This change, however, does not affect the deviance much as in that plot participant 162 does not stand out. We verified the solution (not shown) and  see no reason to delete this participant from the data. 

\begin{sidewaysfigure}[htbp]
\begin{center}
\includegraphics[width = 0.8\textwidth]{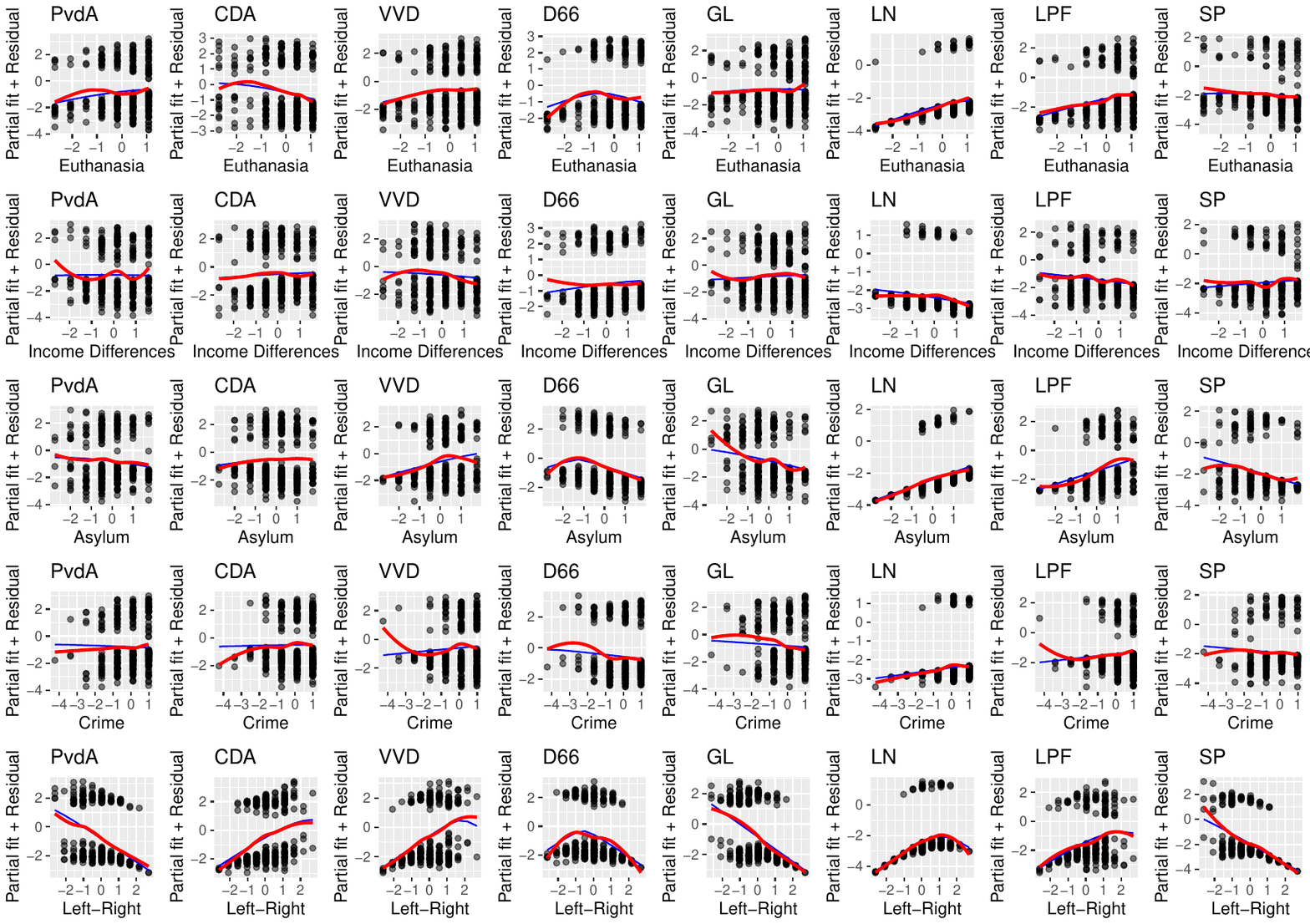}
\caption{Component plus residual plots}
\label{fig:compres}
\end{center}
\end{sidewaysfigure}

In Figure \ref{fig:compres}, we show the component plus residual plots for our mapping. Although there seems to be some misfit for certain predictor response relationships, e.g., income differences for PvdA or crime for VVD, overall there seems to be no reason for large concern. Note that if we would change the functional form of income differences by for example also including a squared effect term, this would not only affect the relationship to PvdA but also the relationship towards all other response variables. 

\newpage 

\newpage

\section*{Acknowledgement}

The Dutch Election data utilized in this manuscript were originally collected for
the Dutch Parliamentary Election Studies 2002 and 2003 by Galen A. Irwin, Joop
J.M. van Holsteyn and Josje M. den Ridder on behalf of the Foundation for Electoral
Research in the Netherlands (Stichting Kiezersonderzoek Nederland, SKON). These
studies have been made possible by grants from Dutch Organization for Scientific
Research (NWO), the Ministry of the Interior and Kingdom Relations (BZK), the
Remote E-Voting Project (Kiezen op Afstand, KOA) of the Ministry of the Interior
and Kingdom Relations (BZK), the Ministry of Health, Welfare and Sports (VWS),
the Social and Cultural Planning Office (SCP), and the Department of Political
Science, Leiden University. The original collectors of the data do not bear any
responsibility for the analyses or interpretations published here. 

We would like to thank the reviewers and guest editors for their constructive remarks on an earlier version of this paper. 

\section*{Declarations}

\paragraph{Funding} No funding was received for conducting this study.
\paragraph{Conflicts of interest/Competing interests} None.
\paragraph{Ethics approval} Not applicable.
\paragraph{Consent to participate} Not applicable.
\paragraph{Consent for publication} Not applicable.
\paragraph{Availability of data and material} The Dutch Election data are publicly available after registration from \texttt{https://easy.dans.knaw.nl/ui/datasets/id/easy-dataset:31979}. 
The Sugiyama data are reported in the paper by Takane (1998) and can be obtained from the corresponding author. 
\paragraph{Code availability} The code for estimation of the supervised and unsupervised maps is implemented in the R-package \texttt{lmap}.
\paragraph{Authors' contributions} MDR developed the supervised and unsupervised maps and derived properties of the maps. MDR with FB derived the algorithm and implemented it in R-code. MDR wrote initial draft of the paper and finalized the manuscript. MDR supervised DW in the data analyses and simulation studies; DW performed the data analyses and simulation studies, commented on the first draft of the paper; FB translated parts of the R-code to C++ code. FB commented on the first draft.

\bibliographystyle{apalike}
\bibliography{melodic.bib}

\end{document}